\pdfoutput=1
\documentclass[12pt,a4paper,leqno]{article}
\usepackage[utf8]{inputenc}
\usepackage{geometry}
\usepackage{xcolor}
\usepackage[round]{natbib}
\usepackage{graphicx}
\usepackage{booktabs}
\usepackage{array}
\usepackage{float}
\usepackage{siunitx}
\usepackage{amssymb,amsbsy,amsmath,bm,mathtools}
\usepackage{setspace}
\usepackage{footmisc}
\usepackage[flushleft]{threeparttable}

\usepackage{caption}
\usepackage{subcaption}
\definecolor{nblue}{HTML}{000660}
\usepackage[colorlinks=true,urlcolor=nblue,linkcolor=nblue,citecolor=nblue]{hyperref}

\newcommand{\Figref}[1]{\hyperref[{#1}]{Figure~\ref*{#1}}}

\begin{document}
\emergencystretch 3em

\title{\textbf{Fueling Inflation Risks: \\
An Oil Shortage Scenario}\thanks{Part of the work by Christiane Baumeister was conducted under the auspices of the Alexandre Lamfalussy Senior Research Fellowship at the Bank for International Settlements whose hospitality is gratefully acknowledged. The views expressed herein are solely those of the authors and
do not necessarily reflect the views of the Federal Reserve Bank of Cleveland, the Federal Reserve System, or the BIS.}}

\author{Christiane Baumeister\thanks{University of
Notre Dame, NBER and CEPR. Email: cjsbaumeister@gmail.com} \and
Todd E.\ Clark\thanks{Johns Hopkins University and Federal Reserve Bank of Cleveland. Email: tclarkresearch@gmail.com} \and
Florian Huber\thanks{University of Salzburg. Email: florian.huber@sbg.ac.at}}

\date{November 2026}
\maketitle
\thispagestyle{empty}

\renewcommand{\baselinestretch}{1.2}
\makeatletter
\g@addto@macro\normalsize{%
  \setlength\abovedisplayskip{6pt plus 2pt minus 2pt}%
  \setlength\abovedisplayshortskip{2pt plus 2pt}%
  \setlength\belowdisplayskip{6pt plus 2pt minus 2pt}%
  \setlength\belowdisplayshortskip{4pt plus 2pt minus 2pt}%
}
\makeatother

\begin{center}
\begin{minipage}{0.9\textwidth}
\noindent\small
\begin{center}\textbf{Abstract}\\\end{center}
Central banks routinely use alternative scenarios to assess risks to the economic outlook, but these are usually generated from linear models that are unable to speak to the likelihood of extreme outcomes. We develop a nonlinear modeling framework that turns a narrative oil market scenario into a fully probabilistic assessment of the dynamics of oil and gasoline prices, aggregate inflation, and household inflation expectations. It combines an expert-based path for a tail event, a nonparametric model of the global oil market, and nonlinear Phillips curves for the energy, food, and core components of PCE inflation. We apply it in real time to quantify energy price and inflation risks of a severe oil shortage caused by the closure of the Strait of Hormuz, modeled as a rapid drawdown of global petroleum inventories, using only information available in February 2026. Our analysis shows that once oil reserves hit rock bottom, oil and gasoline prices rise sharply, with risks tilted to the upside. Headline PCE inflation peaks near 4.8\% in late 2026, and households expect inflation to rise by about 3 percentage points. Two years out, inflation is still highly likely to exceed the Fed's 2\% target.

\bigskip

\end{minipage}
\end{center}

\setcounter{page}{0}
\normalsize\newpage

\section{Introduction}\label{sec:intro}
The closure of the Strait of Hormuz in the wake of the US attack on Iran caused an unprecedented disruption of global oil flows. To offset the resulting supply shortfall, the International Energy Agency (IEA) swiftly acted to coordinate a release of 400 million barrels of crude oil and refined products from emergency stockpiles among its member countries, the largest intervention ever \citep[see][for an analysis of previous episodes]{Baumeister2023Pandemic}. Together with record drawdowns from commercial inventories, supplies from storage initially kept a lid on oil prices. But the inventory buffer is finite. Once this shock absorber is exhausted, any remaining oil deficit has to be balanced by price adjustments, and the risk of extreme price increases rises sharply \citep{DeatonLaroque1992}. A prolonged blockade of tanker traffic through the Strait therefore also poses an important upside risk to the inflation outlook. This raises the questions of how high oil prices, gasoline prices, and inflation could go if global petroleum inventories are drained to critical levels, and how likely such outcomes are. We address these questions with a scenario analysis that forms the basis for our real-time risk assessment.

Monetary policymakers around the world have long emphasized the need to inform policy decisions made under uncertainty with an analysis of alternative scenarios to assess risks around the baseline economic outlook \citep[see, e.g.,][]{BordoetalNBERw27369, garga2025monetary, cairo2025risk}.\footnote{For example, \cite{bernanke2024forecasting} recommends in his review of forecasting at the Bank of England that ``the central forecast should be regularly augmented by alternative scenarios" to ``help to assess the effects and costs of possible risks to the outlook arising from unexpected changes in exogenous variables."} As documented in \cite{herbst2026alternative}, the Federal Reserve Board has historically focused on several broad classes of scenarios or risks, one of which is associated with oil markets.\footnote{From our own tabulations of publicly available Greenbook/Tealbook analyses of alternative scenarios for the period from 2005 through 2019, oil price scenarios featured in six out of the 15 years (2008, 2011, 2012, 2013, 2017, and 2018), underscoring the salience of oil price risks in policy deliberations.} These scenarios, however, are typically produced with linear models or derived from market-based information and presented as point paths.\footnote{A case in point is the study by \citet{Kilianetal2026} that assesses the impact of the Iran war on US inflation using a linear vector autoregressive model conditional on a scenario-implied oil price path from a calibrated DSGE model without taking uncertainty, risk, and nonlinearities into account.} They are silent on how likely extreme outcomes are under a given scenario, even though it is precisely these tail risks that policymakers care most about in a situation of heightened uncertainty. In the current episode, for example, FOMC participants noted that upside risks to inflation remained elevated, in part because of supply shocks in energy.\footnote{See the minutes of the June and July 2026 FOMC meetings.}

We develop a flexible and transparent framework that maps an oil market scenario into probabilities of outcomes for gasoline prices, various measures of inflation, and household inflation expectations, enabling us to quantify tail risks for the inflation outlook. Our framework consists of three components. The first is an expert assessment of possibly severe developments in global oil markets that draws on a wide range of information to construct a path for global petroleum inventories over roughly the next two years. The resulting scenario postulates that both emergency reserves and commercial stockpiles are drawn down at record speed to their all-time low by the end of June 2026, remain at that operational minimum for one year, and then gradually recover to half of the pre-war level.

The second component is the nonparametric model of \cite{BaumeisterHuberMarcellino2024RiskyOil}, which combines a linear vector autoregression (VAR) with Bayesian additive regression trees \citep[BART,][]{Chipmanetal2010} and factor stochastic volatility. This model was specifically designed to capture tail risks in global oil markets, including the risk of stockouts, and it has been shown to produce accurate tail forecasts of the real price of oil. We use this nonlinear model to generate predictive distributions of the real price of oil and other oil market fundamentals conditional on the predetermined inventory path and translate each simulated value for the oil price into US retail gasoline prices, nationally and for 474 metropolitan areas, using the rule of thumb of \citet{Hamilton2012}. We examine the conditional dynamics of gasoline prices because they are among the most visible and frequently observed prices for consumers, greatly influencing the formation of household inflation expectations, which in turn affect future inflation.

The third component uses direct multi-step Phillips curves for the energy, food, and core components of personal consumption expenditure (PCE) price index, and analogous models for household inflation expectations, all specified with BART to capture nonlinearities and tail risks \citep[see][]{clark2023tail, ClarkHuberKoopMarcellino2024Inflation}. Feeding the scenario-implied oil price draws through these models yields predictive distributions for inflation measures that we use to compute the probability of exceeding policy-relevant thresholds and the expected excess over a threshold if it is crossed, in the spirit of the risk measures of \citet{BaumeisteretalRESTAT2022}.

It is important to keep in mind that our approach is not meant to produce an out-of-sample forecast but to provide a hypothetical analysis to assess worst-case outcomes and facilitate central bank risk management. Accordingly, our scenario analysis is deliberately designed to study the consequences of a low-probability but potentially highly damaging event that unfolds over several months, triggered by a policy intervention. To illustrate the usefulness of our framework for real-time risk assessment, we condition the analysis on information available in February 2026 and do not update the model with data released after the onset of the Middle East conflict.

Our methodological contribution is to show how a narrative scenario can be turned into fully probabilistic statements about oil price and inflation risks. Relying on models with nonlinear dynamics and time-varying volatility is critical for this purpose, because the scenario can change not only the location of the predictive distribution but also its shape. Existing approaches for scenario analysis \citep[e.g.,][]{WaggonerZha1999, BaumeisterKilian2014IMFER, AntolinDiazetal2021} deliver conditional forecasts from linear Gaussian models that are unable to account for tail events and asymmetries. Models that forecast and quantify tail risks to the macroeconomic outlook \citep[e.g.,][]{adrian2019vulnerable, clark2023tail, LopezSalidoLoria} do so unconditionally, while we show how risk assessments can be made conditional on a specific, policy-relevant scenario.

Our analysis shows that the rapid release of inventories modeled after the IEA's prompt policy intervention had a stabilizing effect on oil prices in the early stages of the Hormuz crisis. But once petroleum stockpiles are drained to historic lows, physical shortages set in motion an adjustment process in the global oil market. The median real price of Brent climbs steadily to \$166 per barrel by mid-2027 and the probability of oil prices exceeding \$200 rises to 28\% at the end of 2027. The scenario-implied Brent price for October 2026 is \$140 compared to a realized price of \$125. The erosion of inventories also exerts substantial upward pressure on refining margins, which rise steeply by more than \$40 per barrel between June and October 2026, in line with recent developments in actual gasoline and diesel spreads. Despite market tightness, oil investment, proxied by oil drilling activity, remains subdued until early 2027, consistent with oil companies prioritizing debt reduction and shareholder payouts over increasing production capacity. The surge in oil prices also induces some demand destruction but the odds favor fuel consumption remaining elevated. The median path for the national average of gasoline prices evolves in lockstep with the Brent price, exceeding its June 2022 record of \$5/gal one year after the onset of the military conflict, with risks tilted to the upside and unevenly spread across geographic locations.

Given the well-documented sensitivity of households to energy prices \citep[e.g.,][]{CoibionGorodnichenko2015, JoKlopack2025, BaumeisterFrankHuberKoop2026}, the surge in prices at the pump raises household inflation expectations by roughly 3 percentage points to a peak of around 9\% in early 2027. The scenario-implied increase in the aggregate masks differences across demographic groups, with some survey participants expecting inflation as low as 6\% and others as high as 11\%, in line with evidence provided by \citet{PfajfarSantoro2008} and \citet{Hajdinietal2024}. The price pressures from the weakening inventory buffer result in a sharp increase in headline PCE inflation over the course of 2026, with the median projection peaking at nearly 4.8\% in December. Excluding food and energy prices, the median projection of core PCE inflation stays at or above 3\% until about mid-2027 with a peak around 4\% in early 2027. Two years after the outbreak of the war, the probability that inflation exceeds the FOMC's longer-run inflation objective of 2\% is 86\% for headline and 97\% for core inflation.

The paper proceeds as follows. Section 2 introduces a tail risk scenario designed to study the implications of the massive drawdown of petroleum inventories for the predictive distributions of the real Brent price and other supply and demand indicators over a two-year horizon. Section 3 maps the conditional densities of oil price forecasts to US retail gasoline prices at the national, state, and metropolitan levels and conducts a risk assessment of exceeding various price thresholds at the pump. Section 4 studies the scenario-induced consequences for inflation dynamics using a nonlinear Phillips curve framework that generates time-specific predictive densities. In particular, we examine the upside risks for core and headline inflation as well as the adjustment of household inflation expectations with a special focus on heterogeneity across demographic groups. Section 5 offers some concluding remarks.

\section{A Tail Risk Scenario of the Iran War}\label{sec:setup}
We conduct a scenario analysis centered on the rapid withdrawal of global petroleum inventories initiated by the International Energy Agency's authorization for a record release of 400 million barrels of crude oil and refined products from emergency stockpiles shortly after the outbreak of the conflict in the Middle East. In \autoref{sec:model}, we briefly describe the nonlinear modeling framework that allows us to produce time-specific forecast densities based on information available as of the end of February 2026. In \autoref{sec:oil}, we present the scenario and describe its implementation, and in \autoref{sec:oildens}, we examine its implied effects on oil market dynamics. While we focus on a particular scenario that loomed large in the early stages of the Iran war, this state-of-the-art model delivers a blueprint for assessing a range of tail risks in oil markets and should be a useful tool for analyzing the implications of hypothetical worst-case outcomes in future energy crises.

\subsection{A Tail Forecasting Model of the Global Oil Market}\label{sec:model}

To model the tail behavior of oil market fundamentals, we rely on the nonparametric model of the global oil market recently proposed by \cite{BaumeisterHuberMarcellino2024RiskyOil}, which takes the following form:\footnote{As shown in \cite{BaumeisterHuberMarcellino2024RiskyOil}, this model produces accurate tail forecasts for the period from 1992M1 to 2019M12. In \autoref{tab:tails2020}, we extend the forecast evaluation to the period 2020M1 to 2026M2 and show that the nonparametric model outperforms two random walk benchmarks and a VAR with stochastic volatility also over this sample.}
\begin{equation}
\bm{y}_{t}=\underbrace{\bm{Ax}_{t}}_{\text{linear part}}
+\underbrace{G(\bm{x}_{t})}_{\text{flexible part}}
+ \hspace{0.3 cm} \bm{\varepsilon }_{t},\qquad
\bm{\varepsilon }_{t}\sim \mathcal{N}(\bm{0},\bm{\Sigma }_{t}),
\end{equation}
where $\boldsymbol{y}_{t}$ denotes a $7$-dimensional vector that contains monthly data on the real price for physical cargo of Brent crude oil, global fuel consumption, OECD petroleum inventories, the global economic conditions (GECON) indicator of \cite{BaumeisteretalRESTAT2022}, a commodity price index of raw industrial materials deflated by US consumer prices, the gasoline-Brent price differential as a proxy for the refining margin, and worldwide rig counts, and $\boldsymbol{x}_{t}=(\boldsymbol{y}_{t-1}^{\prime },...,%
\boldsymbol{y}_{t-p}^{\prime })^{\prime }$ includes $p=12$ lags of $%
\boldsymbol{y}_{t}$.\footnote{%
All variables enter the model in logs except for GECON and the gasoline-Brent spread. The model is estimated with real-time data based on the February 2026 vintage with the first observation in 1975M1. Given delays in the release of some variables, we apply simple nowcasting rules based on the time-series properties of the data to fill in gaps at the end of the sample. Appendix A lists the data and their sources. More details on the variable definitions, associated risks, and nowcasting rules can be found in \cite{BaumeisterHuberMarcellino2024RiskyOil}.} The relationship between the
outcome and predictor variables consists of two components. The term $\bm A \bm x_t$ captures linear dynamics. This is the part of the model which explains most of the variation in $\bm y_t$ in normal times. The second term $G(\bm x_t)$ captures nonlinear features between $\bm y_t$ and $\bm x_t$ which are pervasive in oil markets, especially in times of crises.  We need not take a stand on where the nonlinearity originates or what form it takes. For instance, if the pass-through from inventories to prices steepens once stocks fall below some historically low threshold, our model learns the shape of this changing relationship from the data rather than relying on a specific functional form imposed a priori. This is exactly the situation we consider in our scenario where the same drawdown of inventories might propagate very differently depending on the current state of the market. It is in such situations that estimating a relationship under the assumption of linearity would be most misleading and that allowing for a flexible conditional mean function pays off the most. Another key feature of the model is that the shock variance $\boldsymbol{\Sigma }_{t}$ is allowed to drift over time, which means that the predictive bands widen automatically in turbulent periods and tighten in calm ones instead of being anchored at a single historical average.\footnote{The time-varying covariance matrix follows the factor stochastic volatility specification of \cite{PittShephard1999} and \cite{AguilarWestJBES2000}; see \cite{BaumeisterHuberMarcellino2024RiskyOil} for details.}

We estimate the unknown function $G(\bm x_t)$ with Bayesian additive regression trees (BART) as in \cite{Chipmanetal2010}. To illustrate how BART works, we consider a single regression tree which is simply a sequence of yes/no questions about the predictors.  For instance, such questions include: ``Were inventories below the historical average last month?'' or ``Were global economic conditions above their long-run average?'' These binary questions are then combined to partition observations into several groups (which are not known and depend on the number of questions asked), and the fitted value for a particular observation is then the average outcome within the group that this observation belongs to. One tree on its own is a coarse step function and typically forecasts poorly. The key idea of BART is to add up many such trees, each kept deliberately shallow by a prior so that it explains only a small fraction of the variation---a ``weak learner'' in machine-learning language. Summing over hundreds of these simple, regularized trees can trace out rich and possibly nonlinear dynamics, while the prior disciplines each tree and guards against overfitting.\footnote{%
Each equation-specific function $g_{i}(\bm{x}_{t})$ in $G(\bm{x}_{t})=(g_{1}(\bm{x}%
_{t}),...,g_{n}(\bm{x}_{t}))^{\prime }$, with $n=7$, is approximated by a sum of $S$ regression trees.  \cite{Chipmanetal2010} and \cite{HUBER202352} find that forecast accuracy improves with the number of trees up to $S\approx 50$ and is roughly flat thereafter; we follow the standard choice of \cite{Chipmanetal2010} and set $S=250$. The remaining prior settings are the defaults of \cite{Chipmanetal2010}: the depth-penalizing tree prior with base and power parameters $(0.95, 2)$, Gaussian terminal-node priors with scale parameter $k=2$, and the inverse chi-squared error-variance prior with $(\nu, q)=(3, 0.9)$.}

Because the procedure is Bayesian, it returns a full posterior predictive distribution rather than a single fitted path, which is precisely what is required to attach probabilities to extreme outcomes. The flexible mean specification allows for asymmetric and state-dependent effects on the predictive distribution, while the time-varying volatility component produces forecast densities that are honest about how uncertain the future is under stress. These features are key to tracing out the entire predictive distribution of oil market outcomes under a specific, deterministic scenario, which is what we turn to next.

\subsection{Scenario Design}\label{sec:oil}
We assume that our oil shortage scenario unfolds over several periods starting at the end
of February 2026. Specifically, the coordinated release of oil from
government-controlled emergency stockpiles is modeled to take place over roughly 120 days.\footnote{%
The timeframe of 120 days for the release of 172 million barrels from the
SPR was projected by the US Department of Energy at the time of the
announcement (see
\url{https://www.energy.gov/articles/united-states-release-172-million-barrels-oil-strategic-petroleum-reserve}). Other IEA member countries anticipated similar trajectories for their deliveries.} We further assume that this policy intervention is accompanied by rapid drawdowns from commercial inventories held as part of
normal business operations by oil producers, refiners, traders, and
distributors.\footnote{The US Energy Information Administration reported in its Weekly Petroleum Status Report that commercial inventories of crude oil and refined products have fallen precipitously and are well below their five-year average. Leaders from top oil companies like Exxon Mobil and Chevron have also explicitly stated that stocks are approaching multi-decade lows last seen in 1983 (see, for example, \url{https://www.politico.com/news/2026/06/04/oil-price-spike-white-house-hormuz-00949435}).} While privately held stocks were within historical fluctuations of the past five years before the outbreak of
the conflict, strategic petroleum reserves (SPR) in the US and other IEA member
countries had been insufficiently replenished after the massive withdrawals
in the wake of the Russian invasion of Ukraine in 2022. For example, the US SPR was only at 58\% of its authorized capacity of 713.5 million barrels in February 2026. Given the
unprecedented pace of destocking, we postulate that the historic floor is reached at the end
of June 2026 and that inventories remain at their operational
minimum for 12 months before gradually recovering to half the pre-war level
over the subsequent six months.\footnote{Operational minimum refers to the lowest possible amount of oil for pipelines, storage tanks, and export terminals to function properly.}
Thereafter, petroleum inventories evolve according to the unrestricted model
dynamics up to the end of the forecast horizon $h$. Thus, the path for our scenario can be summarized as follows:
\begin{equation}
 \mathcal{I}_{T+h} \;=\; \begin{cases}
  I_T + \tfrac{h}{4}\big(I^{\text{MIN}}- I_T\big), & h=1,\dots,4,\\[2pt]
 I^{\text{MIN}},                                          & h=5,\dots,16,\\[2pt]
 I^{\text{MIN}} + \tfrac{h-16}{6}\big(0.5*(I_T-I^{\text{MIN}}\big)), & h=17,\dots,22,\\[2pt]
 \text{unrestricted},                                          & h=23,\dots,
\end{cases}
\end{equation}
where $I_T$ denotes the inventory level at the end of February 2026 and $I^{\text{MIN}}$ the all-time low in February 1977.

The idea underlying this design is that even after the disruptions of traffic through the Strait of Hormuz are fully resolved, the physical market remains extremely tight because it will take considerable time for the complex physical supply chain dynamics to normalize even if a permanent agreement between the US and Iran is reached.\footnote{In an interview with Reuters on August 4, 2026, Aramco CEO Amin Nasser expressed the view that ``if the Strait were to open today, it would take up to 18 months at an average rate of 2.1 million barrels a day to replenish depleted inventories." It is important to note that oil reserves fill about six times slower than they drain. Refilling stocks also requires obtaining oil of a particular grade since mixing crude of different sulfur content can create operational problems.} Tanker traffic is
likely to pick up only gradually given that sea mines need to be removed
from shipping lanes, displaced ships need to be rerouted from elsewhere in
the world, and confidence that the passage of vessels is safe and can happen
without any sort of interference needs to be rebuilt. While it should not
take too long to get the vessels that have been stranded in the Persian Gulf
for months ready to sail, the logistics of this operation are complex,
including the procurement of insurance coverage. It is also important to
keep in mind that oil shipments will reach their destinations only after
several weeks. Thus, once inventories are drained to critical levels,
physical shortages will start to bite, which calls for channeling any
supplies to current consumption rather than storage, prolonging the
inventory deficit. For example, the typical uptick in fuel demand due to the summer travel season is likely to absorb large amounts of available oil.

Other unknowns in the normalization process of export flows are the willingness of shipping companies to reenter the Gulf to load again and the willingness of insurance companies to provide coverage at reasonable rates. The former is key for emptying the region's brimming storage tanks to enable producers to start pumping again and the latter is key for keeping transit safe and affordable given the elevated risks of Iran continuing to control the waterway, including the repeated threat of charging tolls.\footnote{A useful precedent is the slow recovery of tanker traffic in the Red Sea in the aftermath of the Houthi attacks in 2023, with the number of ships passing through still 50\% lower than before the naval hostilities.} It stands to reason that shipping traffic is not governed by physical access alone, but also by risk perceptions and trust, which have the potential to permanently alter shipping behavior and reconfigure global maritime logistics. It will also take a while to ramp up oil production and restore refining activity in the Gulf region to pre-war levels. Not only is it difficult to restart oil wells that have been shut in for an extended period, but the damage from attacks on upstream and downstream energy infrastructure also needs to be repaired, causing additional delays that constrain global oil flows. We assume that all these factors contribute to keeping inventories at their operational lows for one year.

The gradual recovery of inventories after June 2027 is in part motivated by
the fact that the US government did not sell out of the SPR but loaned oil
under the provision to return the same volume to the facility plus a
substantial premium in the form of additional barrels at a later date.%
\footnote{%
In total, the US expects companies to repay approximately 200 million
barrels including the premium, versus the 172 million loaned out, by 2028.}
By that time, we expect the market to have returned to normal operations, which will allow for the severely curtailed inventories to be replenished. In addition to refilling existing facilities, import-dependent countries are likely to establish or expand strategic petroleum reserves and build additional fuel storage capacity to secure supplies and mitigate the impact of future disruptions, which supports a partial rebound of global stocks to 50\% of the amount recorded before the onset of the conflict over a period of six months. Thus, our scenario incorporates important demand-side
aspects.

\subsection{Conditional Predictive Densities of Oil Market Fundamentals}\label{sec:oildens}

Panel (a) of Figure 1 displays the deterministic path for petroleum
inventories laid out in equation (2) with the star indicating the storage level before the outbreak of the war. The multi-step-ahead predictive distributions for the other
oil market fundamentals $\bm y_{T+h}$ are conditioned on this path as follows:%
\begin{equation*}
p(\bm y_{T+h}|\bm \Omega_{1:T},\mathcal{I}_{T+h})
\end{equation*}%
where $\bm \Omega_{1:T}$ denotes the information set up to $T=2026$M$2$. We implement the conditioning by simulation. For each posterior draw, we iterate the model forward one month at a time, draw the shocks from their predictive distribution, and replace the simulated value of inventories with the value implied by the scenario path. The remaining variables respond to the inventory path through the lagged dynamics of both the linear and the nonlinear components of the model. This not only allows us to trace out the dynamic evolution of the most likely path of the model variables implied by our scenario but also nonlinear changes in the distribution of possible outcomes over the forecast horizon with direct implications for risk assessment. It is important to stress that these forecast densities adapt continuously as the model in equation (1) produces different combinations of covariates reflecting prevailing market conditions which might induce different tree allocations.

\begin{figure}[t]
\caption{\centering Scenario-Implied Fan Charts for Oil Market Fundamentals}
\centering
  \includegraphics[width=0.98\textwidth]{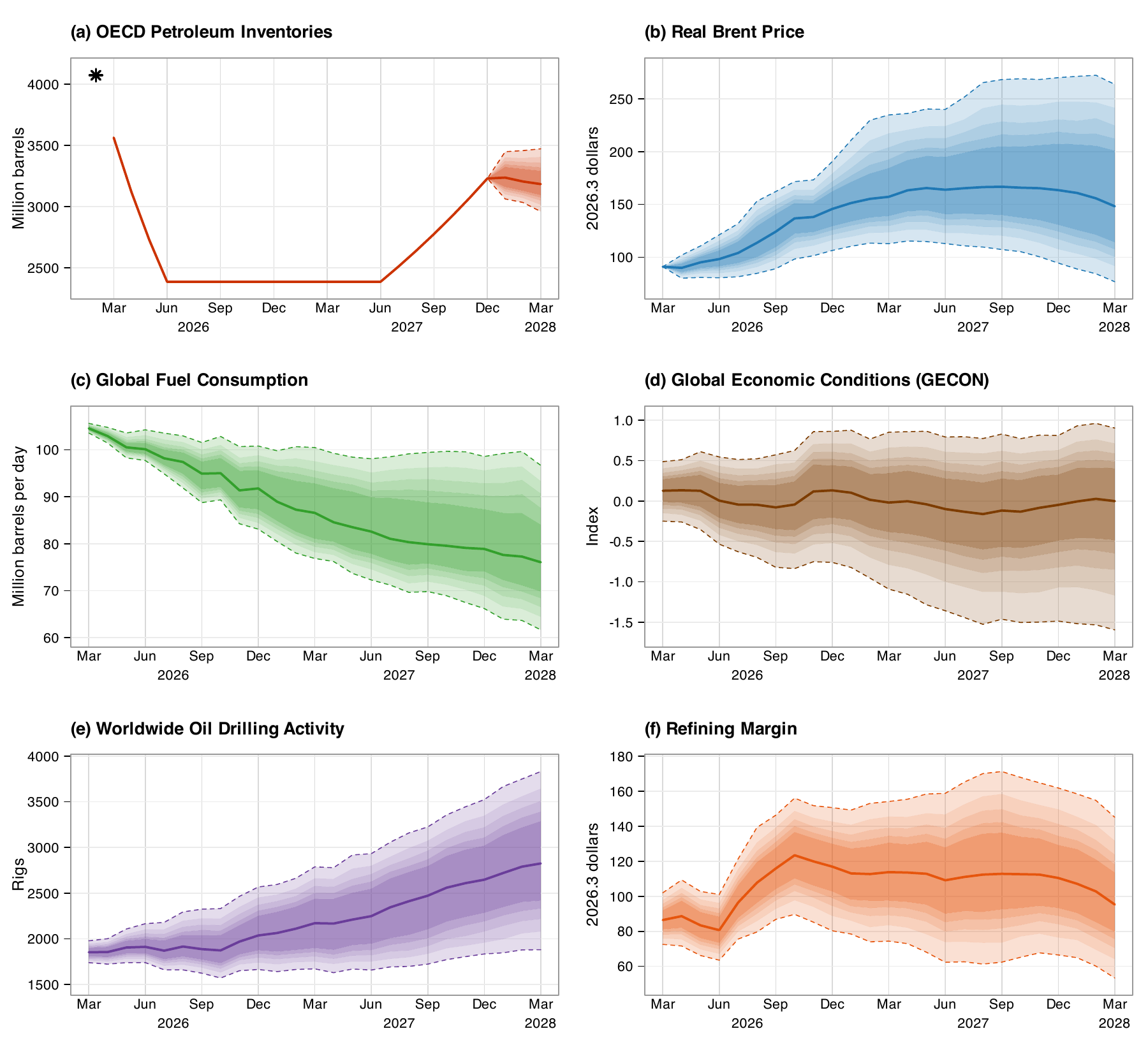}
  \caption*{\footnotesize \textbf{Notes}: The star in panel~(a) marks the February 2026 inventory level. The solid line is the median of the conditional posterior predictive density and the increasingly lighter shaded areas correspond to 50-60-70-80-90\% probability bands.}
  \label{fig:scenario}
  \vspace{-.3cm}
\end{figure}

Panel (b) shows the conditional posterior predictive densities for the real
Brent price expressed in March 2026 dollars in the form of a fan chart with the
solid line being the median path.\footnote{%
Specifically, the price for March 2026 corresponds to the observed oil price
on the day of the IEA announcement. This is the only information from after February 2026 that enters the model-based part of our analysis.} While the swift draining of reserves
implies a monthly oil price of just below \$100 up until June 2026 with
limited upside and downside risks, prices rise steeply once the inventory
buffer is exhausted,\ approaching \$150 by year end and peaking near \$166 in mid-2027. Oil prices are expected to remain above \$150 until the end
of the forecast horizon. With inventories at historic lows for an extended
period, the loss of shock absorbers entails greater risks. This is reflected
in the predictive distribution becoming more skewed to the upside, which
points to a considerable increase in tail risks with the probability of the
oil price exceeding \$200 steadily rising from {8\%} to {28\%} during 2027. Even
after a return to more normal market conditions, the additional demand
deriving from rebuilding depleted emergency reserves and commercial stocks
as well as filling planned and expanded storage tanks will likely put a
floor under oil prices, keeping them higher for longer.

The gradual rise of oil prices rather than an abrupt jump upon stocks
hitting rock bottom is likely the result of physical shortages suppressing
demand. Panel (c) shows that our scenario implies a reduction in global fuel
consumption even in the early stages when lost barrels from the Persian Gulf
are replaced with inventory drawdowns. While the cutbacks are small
initially due to the low short-run price elasticity of oil demand, the
decrease in fuel consumption accelerates over time as households and firms
switch to alternative sources of energy such as electric vehicles for
transportation. What is notable though is that the predictive distributions
put more weight on fuel consumption staying elevated throughout with a long
upper tail. For example, the probability of fuel consumption never falling below 90 million barrels per day over the forecast horizon is {about 12\%},\
indicating that demand destruction is far from certain. Interestingly,
global economic conditions are not expected to be much affected, as can be gleaned from panel (d). While there is a modest slowdown as government and commercial stocks run dry, the scenario-implied median path is overall rather stable. However, the conditional predictive densities imply important asymmetries with material risks to the downside mounting as higher oil prices persist.

Panel (e) suggests that the oil industry is unlikely to drill for more oil and expand production to make up for lost supplies, given that the implied path for worldwide rig counts is essentially flat for most of 2026 due to the uncertainty about the longer-term market outlook. Once there is evidence that oil prices might hold at higher levels, which makes it profitable for oil companies to invest in productive capacity, global drilling activity starts to pick up and rises steadily with no sign of asymmetries in the
predictive distributions. More immediate gains are expected to accrue to refiners the moment inventories reach their all-time low. Panel (f) shows that our scenario implies a steep increase in refining margins, which is in line with market tightness deriving from constrained supplies and surging demand resulting in higher profits for refiners. Prolonged upward pressure on Brent prices due to major storage-driven demand increases in the second half of 2027 keeps refining margins at record highs with risks evenly distributed.

Overall, our scenario analysis paints a comprehensive picture of tail risk dynamics in oil market fundamentals as a result of aggressively driving down global petroleum inventories to fill the supply gap due to the extended closure of the Strait of Hormuz. Once that buffer is exhausted, prolonged physical shortages exert upward pressure on oil prices with risks tilted to the upside, which are likely to spill over to other prices, fueling inflation risks more broadly.

\section{Price Risks at the Pump and Their Geography}\label{sec:gas}
The most immediate impact of rising oil prices manifests itself at the gas pump given that crude oil accounts for roughly half of the total cost of producing gasoline,  implying that retail gasoline prices tend to move in lockstep with oil prices.\footnote{\cite{BaumeisterKilian2016BPEA} document that the pass-through from crude oil prices to retail fuel prices is symmetric with a near one-for-one relationship in the long run.} Since gasoline prices are among the most visible and frequently observed prices for consumers, they play an outsized role in influencing the formation of household inflation expectations which in turn affect future inflation \citep[e.g.,][]{CoibionGorodnichenko2015, Baumeister2023Pandemic}. Another channel through which higher fuel prices work their way into inflation is by increasing firms' transportation costs.  It is therefore of interest to quantify how the projected path for petroleum inventories translates into future price developments at the pump.

Exploiting the historical relationship between oil and gasoline prices, we analyze the scenario-based implications for retail fuel prices by mapping each posterior draw from the predictive distribution of the Brent price to the price of regular gasoline using a transparent rule of thumb. According to \cite{Hamilton2012}, the average US retail gasoline price moves by about 2.5 cents per gallon for every one-dollar change in the price of a barrel of crude oil.\footnote{In Appendix B, we cross-validate this relationship. \cite{Hamilton2012} also shows that the Brent price is the relevant reference for pricing gasoline in the US.} Thus, given a reference pair
$(\bar{p}^{gas},\bar{p}^{oil})$ of gasoline and oil prices, the implied gasoline price at horizon $h$ is
\begin{equation}
p^{gas,d}_{h}=\bar{p}^{gas}+0.025\,\big(p^{oil,d}_{h}-\bar{p}^{oil}\big),
\label{eq:rot}
\end{equation}
where $p^{oil,d}_{h}$ is the Brent price for posterior draw $d$. Applying \eqref{eq:rot} draw by draw yields the implied predictive distribution for gasoline prices. Based on this, we construct two complementary risk measures that allow us to assess the likelihood of retail gasoline prices surpassing a certain threshold. Specifically, we consider the marginal exceedance probability $P^M(p^{gas}_{h}>\tau)$ at horizon $h$, which is estimated as the share of posterior draws above a value $\tau$. We also consider the joint exceedance probability over a window of length $H$, $P^{J}\!\big(\max_{h\le H} p^{gas}_{h}>\tau\big)$, which measures the probability that gasoline breaches $\tau$ in at least one month within the first $H$ horizons. Being computed from each draw's path maximum, it preserves the dependence across horizons and is therefore weakly larger than any single-horizon probability. Finally, we report the expected excess over a threshold conditional on exceeding it, $E[p^{gas}_h-\tau\mid p^{gas}_h>\tau]$, and its probability-weighted counterpart.

\subsection{Aggregate Tail Risks}
\begin{figure}[t]
\caption{\centering Scenario-Implied Fan Chart for US Retail Gasoline Prices and Implied Risks}
  \centering
  \includegraphics[width=0.95\textwidth]{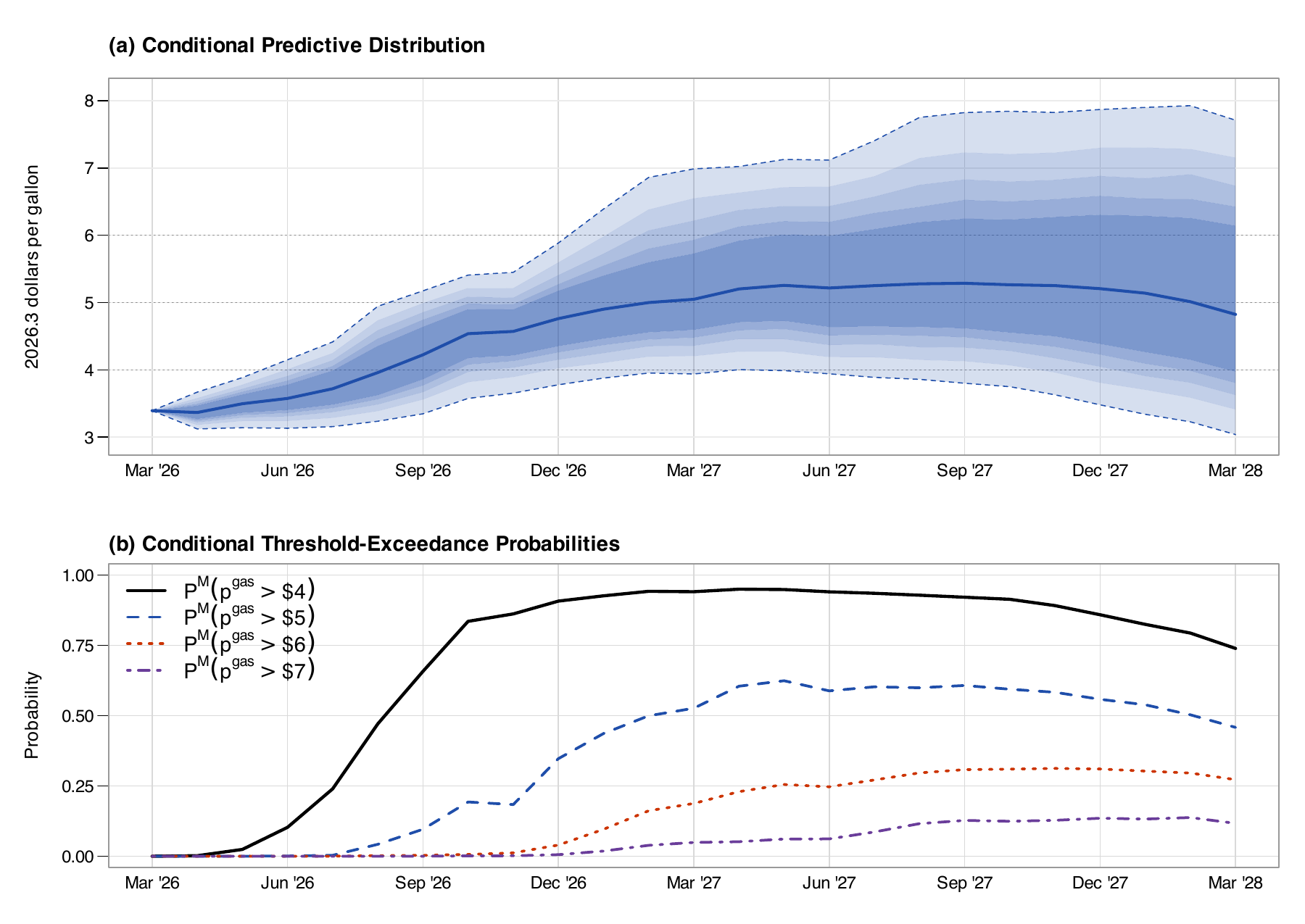}
  \caption*{\footnotesize \textbf{Notes}: The solid line in panel~(a) is the median of the conditional posterior predictive density and the increasingly lighter shaded areas correspond to 50-60-70-80-90\% probability bands.}
  \label{fig:gas}
\end{figure}

Before the outbreak of the Iran war, the average US retail price of regular motor gasoline in February 2026 was \$2.89/gal according to the American Automobile Association (AAA) and the average Brent spot price reported by the EIA was $\$70.89/\text{bbl}$. Using this reference pair in \eqref{eq:rot}, we obtain the conditional predictive distribution of gasoline prices expressed in March 2026 dollars in panel (a) of \Figref{fig:gas}. As can be seen, petroleum stockpiles initially act as a highly effective shock absorber, keeping all possible outcomes for gas prices in the \$3--\$4 range. But once inventories are drained to historical lows, the median path steadily climbs to \$5/gal within less than a year and persistently stays there as global petroleum shortages keep refining markets tight, consistent with the elevated refining margins in panel (f) of \Figref{fig:scenario}. To put these numbers in perspective, during the three years prior to the outbreak of the Iran war, national retail prices of regular gasoline had been fluctuating between \$2.79/gal (recorded in January 2026) and \$3.88/gal (recorded in September 2023).

While projected gas price risks are symmetric until year-end 2026, they become increasingly skewed to the upside with considerable probability attached to tail outcomes. Panel (b) shows the evolution of the probability of exceeding increasingly severe price thresholds of \$4, \$5, \$6, and \$7 per gallon.  Under our oil shortage scenario, the probability that the national average for a gallon of regular gasoline exceeds the \$4 threshold lies between 86\% and 95\% in every month of 2027. One year after the onset of the conflict, there is a better-than-even chance that fuel prices surpass \$5/gal, which was the nominal all-time high reached in June 2022 in the aftermath of Russia's invasion of Ukraine. Prices above \$6 per gallon are not remote either. Their probability rises to about 30\% in the second half of 2027, which would substantially intensify the pain at the pump. Even \$7/gal is not a zero-probability event and constitutes a serious tail risk for drivers. Overall, risks that fuel costs breach various psychological barriers remain elevated as long as the conflict drags on and the supply squeeze persists.

\subsection{Spatial Heterogeneity}
It is well known that retail prices of regular gasoline differ greatly across locations in the US. According to AAA's four-week average pump price as of February 2026, the most expensive gasoline markets were California (\$4.54), Hawaii (\$4.40), and Washington (\$4.15), while the least expensive were Oklahoma (\$2.35) as well as Arkansas, Mississippi, and Kansas (\$2.49 each). This geographical price dispersion arises from several factors. Some states apply higher excise taxes and environmental fees and enforce stricter cleaner-burning fuel blend requirements. Regional market power and distribution margins, limited interstate pipeline connections, and insufficient in-state refining capacity that forces expensive imports also contribute to price differentials.

\begin{figure}[tbp]
\caption{\centering Conditional City-Level Gasoline Price Forecast for November 2026}
  \centering
  \includegraphics[scale=0.65, trim={1cm 2cm 0 2cm},clip]{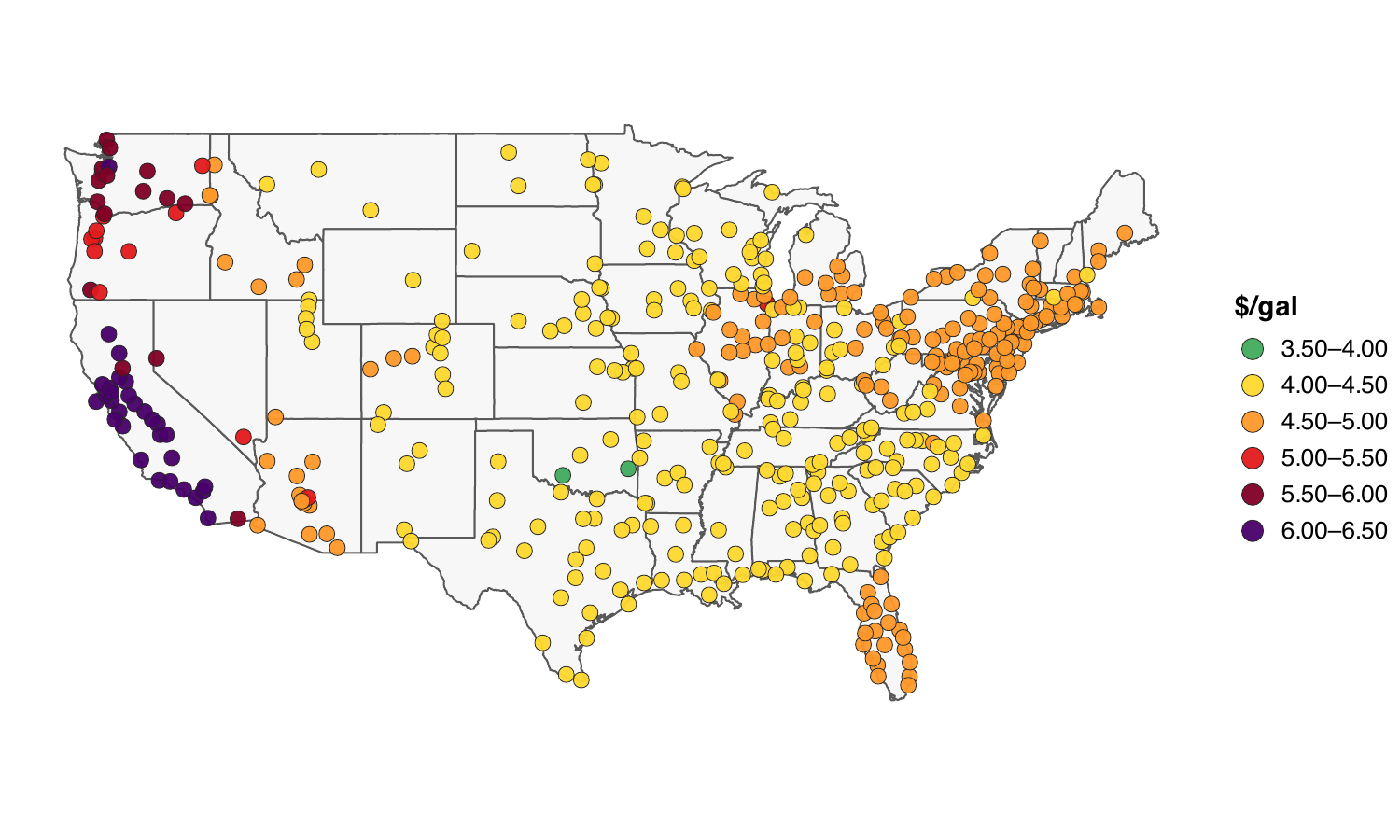}
  \caption*{\footnotesize \textbf{Notes}: Each dot represents the scenario-implied median gasoline price forecast nine months ahead for one AAA metropolitan area in the lower 48 states.}
  \label{fig:geo}
\end{figure}

To analyze the spatial pattern of pump-price risk implied by our inventory scenario, we apply the same rule of thumb at the metropolitan level where $\bar{p}^{gas}$ in the reference pair is now replaced with each geographical unit's average retail price in February 2026.\footnote{AAA collects daily prices for regular gasoline from around 150,000 individual gas stations nationwide and then aggregates over station-level price data in metropolitan areas to obtain city-level and state-level averages.} Since we hold the slope fixed, persistent cross-location differences in baseline prices carry over to the scenario-implied gas price trajectories. While a few states (e.g., Indiana, Georgia, Utah, Illinois, and Kentucky) have enacted temporary gas tax suspensions, reductions, or rate freezes in response to the war to provide some relief at the pump, most US states have kept their fuel taxes in place.

\Figref{fig:geo} shows the conditional median gasoline price forecast across metropolitan areas for November 2026. It highlights the geographical heterogeneity of scenario-implied price developments nine months out. The majority of metro areas cluster around the national average of about \$4.60/gal, with cities in most Midwestern and Southern states sitting below and cities in most Northeastern states just above that price level. On the West Coast, gasoline prices are expected to surpass \$5/gal with almost all cities in California even topping \$6/gal.

\begin{table}[p]
\centering
\caption{\centering Real-Time Risk Assessment of State-Level Prices at the Pump \\ for November 2026 as of February 2026}
\label{tab:gasrisk}
\begin{threeparttable}
\scriptsize
\renewcommand{\arraystretch}{0.92}
\setlength{\tabcolsep}{5pt}
\begin{tabular}{lr@{\hspace{16pt}}cccccc}
\toprule
 & Median & \multicolumn{3}{c}{$\tau=\$4$/gallon} & \multicolumn{3}{c}{$\tau=\$5$/gallon} \\
\cmidrule(lr){3-5}\cmidrule(lr){6-8}
State & (\$/gal) & $P^{M}(p^{gas}>\tau)$ & \begin{tabular}[b]{@{}c@{}}$E[p^{gas}-\tau\mid$\\$p^{gas}>\tau]$\end{tabular} & $P^{M}\times E$ & $P^{M}(p^{gas}>\tau)$ & \begin{tabular}[b]{@{}c@{}}$E[p^{gas}-\tau\mid$\\$p^{gas}>\tau]$\end{tabular} & $P^{M}\times E$ \\
\midrule
California & 6.22 & 1.00 & 2.21 & 2.21 & 0.98 & 1.24 & 1.22 \\
Hawaii & 6.08 & 1.00 & 2.07 & 2.07 & 0.97 & 1.12 & 1.08 \\
Washington & 5.83 & 1.00 & 1.83 & 1.82 & 0.94 & 0.90 & 0.85 \\
Oregon & 5.38 & 0.99 & 1.39 & 1.37 & 0.76 & 0.59 & 0.45 \\
Nevada & 5.29 & 0.98 & 1.31 & 1.29 & 0.71 & 0.55 & 0.39 \\
Alaska & 5.19 & 0.98 & 1.22 & 1.19 & 0.64 & 0.50 & 0.32 \\
Arizona & 4.84 & 0.94 & 0.91 & 0.85 & 0.38 & 0.38 & 0.14 \\
Pennsylvania & 4.80 & 0.93 & 0.87 & 0.81 & 0.34 & 0.37 & 0.13 \\
DC & 4.78 & 0.93 & 0.86 & 0.80 & 0.33 & 0.36 & 0.12 \\
Illinois & 4.69 & 0.90 & 0.79 & 0.71 & 0.26 & 0.36 & 0.09 \\
Vermont & 4.68 & 0.90 & 0.78 & 0.70 & 0.25 & 0.36 & 0.09 \\
New York & 4.68 & 0.90 & 0.78 & 0.70 & 0.25 & 0.36 & 0.09 \\
Maryland & 4.64 & 0.89 & 0.75 & 0.66 & 0.22 & 0.36 & 0.08 \\
Idaho & 4.62 & 0.88 & 0.74 & 0.65 & 0.22 & 0.36 & 0.08 \\
Michigan & 4.60 & 0.87 & 0.73 & 0.63 & 0.20 & 0.36 & 0.07 \\
New Jersey & 4.59 & 0.87 & 0.72 & 0.62 & 0.19 & 0.36 & 0.07 \\
Maine & 4.59 & 0.87 & 0.71 & 0.62 & 0.19 & 0.36 & 0.07 \\
Massachusetts & 4.58 & 0.87 & 0.71 & 0.62 & 0.19 & 0.36 & 0.07 \\
Delaware & 4.58 & 0.87 & 0.71 & 0.62 & 0.19 & 0.36 & 0.07 \\
Connecticut & 4.58 & 0.87 & 0.71 & 0.62 & 0.19 & 0.36 & 0.07 \\
Florida & 4.57 & 0.86 & 0.71 & 0.61 & 0.19 & 0.36 & 0.07 \\
Rhode Island & 4.55 & 0.85 & 0.69 & 0.59 & 0.17 & 0.37 & 0.06 \\
West Virginia & 4.53 & 0.85 & 0.68 & 0.57 & 0.16 & 0.37 & 0.06 \\
Virginia & 4.52 & 0.84 & 0.67 & 0.57 & 0.16 & 0.37 & 0.06 \\
New Hampshire & 4.52 & 0.84 & 0.67 & 0.57 & 0.16 & 0.37 & 0.06 \\
Ohio & 4.49 & 0.83 & 0.66 & 0.54 & 0.15 & 0.37 & 0.05 \\
Indiana & 4.47 & 0.82 & 0.64 & 0.52 & 0.14 & 0.37 & 0.05 \\
Colorado & 4.45 & 0.80 & 0.63 & 0.50 & 0.13 & 0.37 & 0.05 \\
Utah & 4.44 & 0.80 & 0.63 & 0.50 & 0.13 & 0.37 & 0.05 \\
Montana & 4.43 & 0.79 & 0.62 & 0.49 & 0.12 & 0.37 & 0.05 \\
North Carolina & 4.41 & 0.78 & 0.61 & 0.47 & 0.11 & 0.37 & 0.04 \\
Georgia & 4.39 & 0.77 & 0.60 & 0.46 & 0.11 & 0.37 & 0.04 \\
Wyoming & 4.37 & 0.75 & 0.59 & 0.44 & 0.10 & 0.38 & 0.04 \\
Minnesota & 4.34 & 0.74 & 0.57 & 0.42 & 0.10 & 0.37 & 0.04 \\
New Mexico & 4.34 & 0.74 & 0.57 & 0.42 & 0.10 & 0.37 & 0.04 \\
South Dakota & 4.31 & 0.72 & 0.56 & 0.40 & 0.09 & 0.37 & 0.03 \\
Nebraska & 4.30 & 0.71 & 0.55 & 0.39 & 0.09 & 0.37 & 0.03 \\
South Carolina & 4.30 & 0.71 & 0.55 & 0.39 & 0.09 & 0.37 & 0.03 \\
Alabama & 4.28 & 0.70 & 0.54 & 0.38 & 0.08 & 0.37 & 0.03 \\
Kentucky & 4.28 & 0.70 & 0.54 & 0.38 & 0.08 & 0.37 & 0.03 \\
Wisconsin & 4.26 & 0.68 & 0.53 & 0.36 & 0.08 & 0.38 & 0.03 \\
North Dakota & 4.24 & 0.68 & 0.52 & 0.35 & 0.07 & 0.38 & 0.03 \\
Tennessee & 4.23 & 0.66 & 0.52 & 0.34 & 0.07 & 0.38 & 0.03 \\
Iowa & 4.22 & 0.66 & 0.51 & 0.34 & 0.07 & 0.38 & 0.03 \\
Texas & 4.21 & 0.65 & 0.51 & 0.33 & 0.07 & 0.38 & 0.03 \\
Missouri & 4.21 & 0.65 & 0.51 & 0.33 & 0.07 & 0.38 & 0.03 \\
Louisiana & 4.19 & 0.64 & 0.50 & 0.32 & 0.06 & 0.38 & 0.02 \\
Mississippi & 4.17 & 0.62 & 0.49 & 0.30 & 0.06 & 0.38 & 0.02 \\
Kansas & 4.16 & 0.62 & 0.49 & 0.30 & 0.06 & 0.38 & 0.02 \\
Arkansas & 4.16 & 0.61 & 0.49 & 0.30 & 0.06 & 0.39 & 0.02 \\
Oklahoma & 4.03 & 0.53 & 0.43 & 0.23 & 0.04 & 0.40 & 0.02 \\
\bottomrule
\end{tabular}
\begin{tablenotes}\footnotesize
\item \textbf{Notes}: $P^{M}(p^{gas}>\tau)$ denotes the marginal probability of exceeding $\tau$, $E[p^{gas}-\tau\mid p^{gas}>\tau]$ denotes the conditional expected excess in \$/gal, and $P^{M}\times E$ is their product. States are ordered from most to least expensive.
\end{tablenotes}
\end{threeparttable}
\end{table}

From a risk perspective, a more telling concern than the median retail price expectations are the possible upside risks associated with the supply crunch as fuel inventories run out, pushing prices at the pump above thresholds to which consumers may be particularly sensitive. We consider gasoline prices rising above \$4/gal and \$5/gal as tail events that strain households' finances. However, what matters to drivers is not only the likelihood of crossing a certain threshold but also how far above the threshold fuel prices are expected to be if that tail event occurs. \autoref{tab:gasrisk} reports the probability-weighted expected excess as well as its two components, the exceedance probability and the expected loss in the tail, together with the median price forecast as the reference point. We quantify these risks at the state level for the nine-month horizon which coincides with the midterm elections in November 2026, when the cost of filling up the tank is likely to weigh on voters' minds. In all US states, the probability of exceeding the \$4 threshold is more than 50\% and a virtually certain outcome for the three most expensive states, whose gasoline prices are expected to fall in the \$5.80-\$6.20 range. The odds of surpassing the \$5 threshold are at least 30\% for the nine top-ranked states, while for roughly one third of states they are below 10\%. While the occurrence of this tail event would add on average 37 cents for the remaining lower-probability states, the weighted expected excess does not exceed 9 cents. Instead, for the high-gas-price states, the probability-weighted expected additional fuel costs amount to between 12 cents and \$1.22. Californian drivers are the most exposed, facing the risk of paying \$6.22/gal at the pump in November 2026, followed by Hawaii and Washington whose probability-weighted expected excess over the \$5 threshold amounts to \$1.08 and 85 cents, respectively.

\begin{figure}[t]
\caption{\centering Geographical Distribution of Pump-Price Risks}
  \centering
  \begin{subfigure}{\textwidth}\centering
  \caption{\centering {Through year-end 2026}}
    \includegraphics[scale=0.55, trim={0.6cm 2cm 0.6cm 1.5cm},clip]{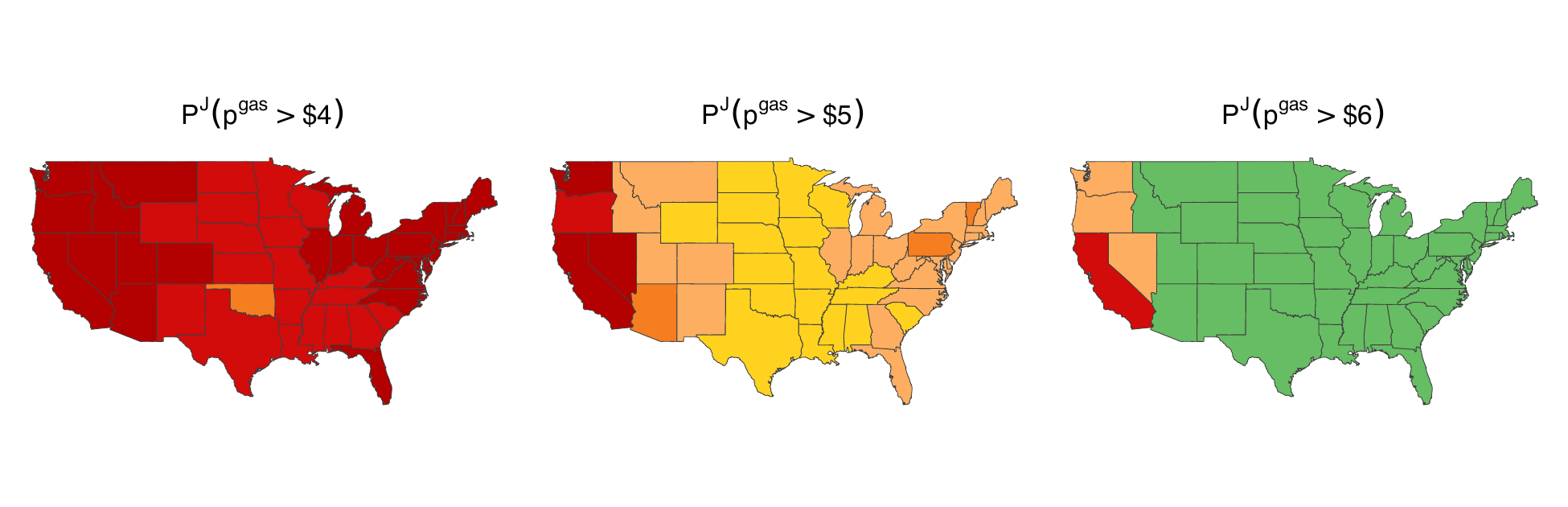}

  \end{subfigure}

  \vspace{0.8em}
  \begin{subfigure}{\textwidth}\centering
  \caption{\centering {Through mid-2027}}
    \includegraphics[scale=0.55, trim={0.5cm 2cm 0 1.5cm},clip]{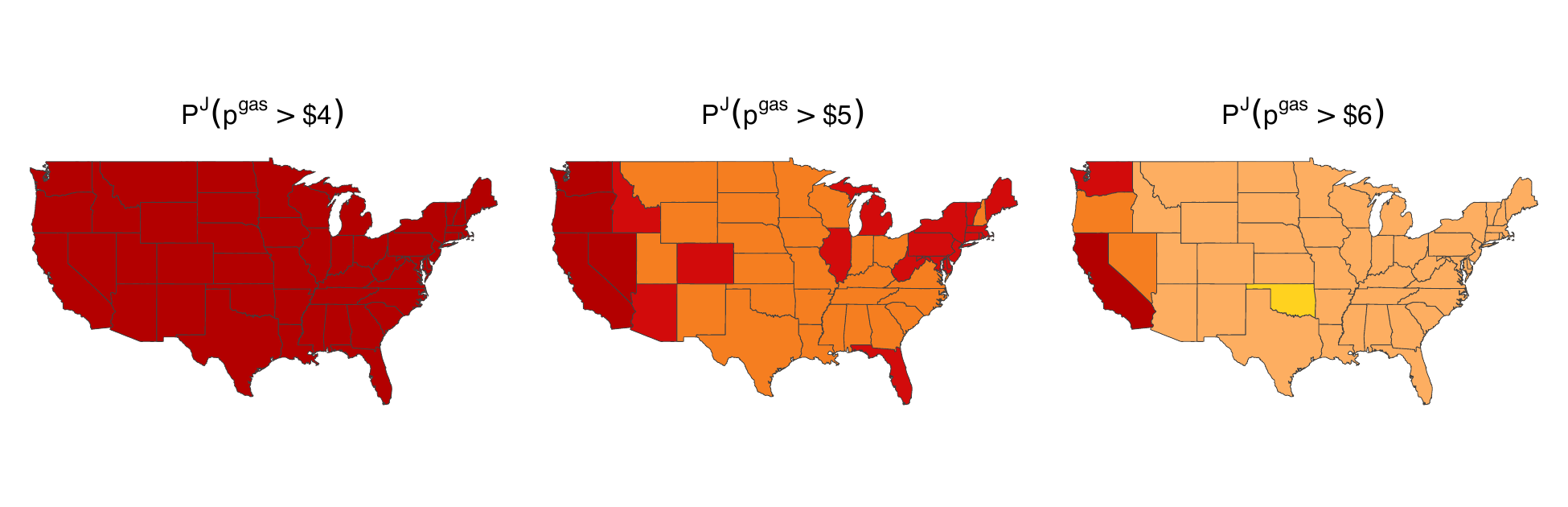}

  \end{subfigure}

  \vspace{2em}
  \includegraphics[width=0.95\textwidth]{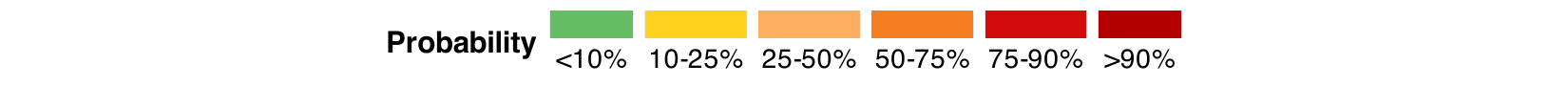} \\
  \vspace{0.4cm}
  \caption*{\footnotesize \textbf{Notes}: While Alaska and Hawaii are part of the analysis, the maps only display the joint exceedance probabilities for the lower 48 states.}
  \label{fig:geoprob}
  \vspace{-.6cm}
\end{figure}

To get a better sense of the risk dispersion across the country further out, \Figref{fig:geoprob} visualizes the joint exceedance probabilities until year-end 2026 and mid-2027 for three pump-price thresholds. The longer the supply constraints bind, the greater the likelihood of tail events at gas stations. Panel (a) shows that by the end of 2026 crossing \$4/gal at some point is a near certainty nationwide, with a median state probability of 92\%. For 10 states the probability of retail gasoline prices rising above \$5 within that timeframe is greater than 50\% with most of the risks concentrated in the West. Prices spiking past \$6/gal by year end remains a tail risk confined to California and Hawaii but the odds of this outcome increase coast-to-coast once the horizon is extended to mid-2027 (panel b), as damaged refineries and missing oil shipments from the Middle East keep fuel markets under pressure. Over this longer window, motor fuel prices in excess of \$4/gal should be taken for granted everywhere; even \$5/gal poses a widespread risk with all states facing at least a 50\% chance of surpassing that threshold, the lowest being 52\% in Oklahoma.

\section{Risks to the Inflation Outlook}\label{sec:inflation}
While adverse developments in global oil markets tend to first show up in domestic gasoline prices, they also pose substantial risks to the outlook for overall inflation given that elevated energy prices raise production costs which are likely to filter through to consumer prices. Additional inflationary pressures might arise from expectational dynamics given that households are highly sensitive to energy price fluctuations \citep[see, e.g.,][]{CoibionGorodnichenko2015, BaumeisterFrankHuberKoop2026}. Armed with the predictive distributions of the real spot price of oil under our shortage scenario, we now examine its implications for inflation risks which matter for macroeconomic outcomes and play a key role in monetary policy decisions.

\subsection{A Nonlinear Phillips Curve}
While most studies on inflation dynamics rely on parametric linear Phillips curve models, a growing body of work considers nonlinear and/or nonparametric alternatives \citep[see][and references therein]{ClarkHuberKoopMarcellino2024Inflation}. One strand of that literature has provided evidence for regime-specific nonlinearities where the slope of the Phillips curve steepens with a stronger economy \citep[see, e.g.,][]{Nalewaik2016, BabbDetmeister2017, Daoetal2024, BenignoEggertsson2025}, whereas other contributions have documented the benefits of nonparametric machine learning methods such as random forests \citep{medeiros2021forecasting, GouletCoulombeJAE2024} and Gaussian processes \citep{ClarkHuberKoopMarcellino2024Inflation} for inflation forecasting.

We follow the latter approach for our analysis and use a nonlinear Phillips curve framework with a direct multi-step BART specification at the monthly frequency which for horizon $h$ takes the form:
\begin{equation}
\pi^{j}_{t+h}=f_{j,h}(\bm z_{t}, \Delta_{12}\log p^{{oil}}_{t+h})+\varepsilon_{t+h}, \qquad
\varepsilon_{t+h}\sim \mathcal{N}(0,\sigma^2).
\label{eq:pcbart}
\end{equation}
The outcome variable $\pi^{j}_{t+h}$ is the 12-month inflation rate in period $t+h$ for $j\in\{\text{energy}, \text{food}, \text{core\}}$, the three components of personal consumption expenditures (PCE). The forecast of headline PCE inflation is constructed bottom-up as the expenditure-share-weighted sum of the individual forecasts for energy, food, and core inflation. Each $f_{j,h}$ is a sum-of-trees BART ensemble which allows the pass-through of a given indicator to a particular measure of inflation to be nonlinear and state-dependent; for example, the effect of oil prices can be stronger when inflation is already elevated \citep[see][for a model with this property]{HardingLindeTrabandt2023}. The set of predictors in $\bm z_{t}$ closely follows the selection of theory-inspired determinants used in the inflation-at-risk analysis of \cite{LopezSalidoLoria} and contains the lagged own inflation rate to capture inflation persistence, the output gap as a measure of slack in the economy, year-ahead household inflation expectations from the Michigan survey, and the excess bond premium of \citet{GilchristZakrajsek:2012:AER} as a proxy for financial conditions. To account for the specifics of the economic environment at the end of February 2026, we also include the geopolitical risk indicator of \citet{CaldaraIacoviello2022} intended to capture impacts of war-related events that go beyond oil prices, and the consumer sentiment index from the Michigan survey to capture shifts in household attitudes.\footnote{Appendix A lists the data and their sources.} Among the predictor variables, only the year-on-year percent change in the real Brent price enters contemporaneously for our scenario analysis.\footnote{The use of the real rather than nominal oil price as a determinant of inflation dynamics follows Phillips curve specifications commonly used in the literature \citep[see, e.g.,][]{Yellen2015, ConflittiLuciani}.} In contrast to the oil market model, we treat the innovations $\varepsilon_{t+h}$ as homoskedastic since \citet{ClarkHuberKoopMarcellino2024Inflation} have shown that once BART captures nonlinearities of unknown form in the mean, adding stochastic volatility does not improve the forecasting performance for inflation further on average.

We estimate each model separately using the Bayesian approach of \cite{Chipmanetal2010} for BART, with data running from January 1978, which is when the Michigan survey starts, to February 2026. Feeding each projected oil price path through the time-specific Phillips curves delivers conditional posterior predictive distributions of inflation outcomes for each horizon that reflect parameter uncertainty and future shocks to inflation, conditional on the oil price draws, and that we use to study the dynamics of inflation tail risks.

\subsection{Inflation Tail Dynamics}
We first describe the scenario-implied predictive distributions of various inflation measures and the associated tail risks, and then compare them with survey-based density forecasts taken before the war.

\subsubsection{Scenario-Implied Inflation at Risk}
\begin{figure}[t]
\caption{\centering Scenario-Implied Fan Charts of Inflation}
\vspace{-.1cm}
  \centering
  \includegraphics[width=0.95\textwidth]{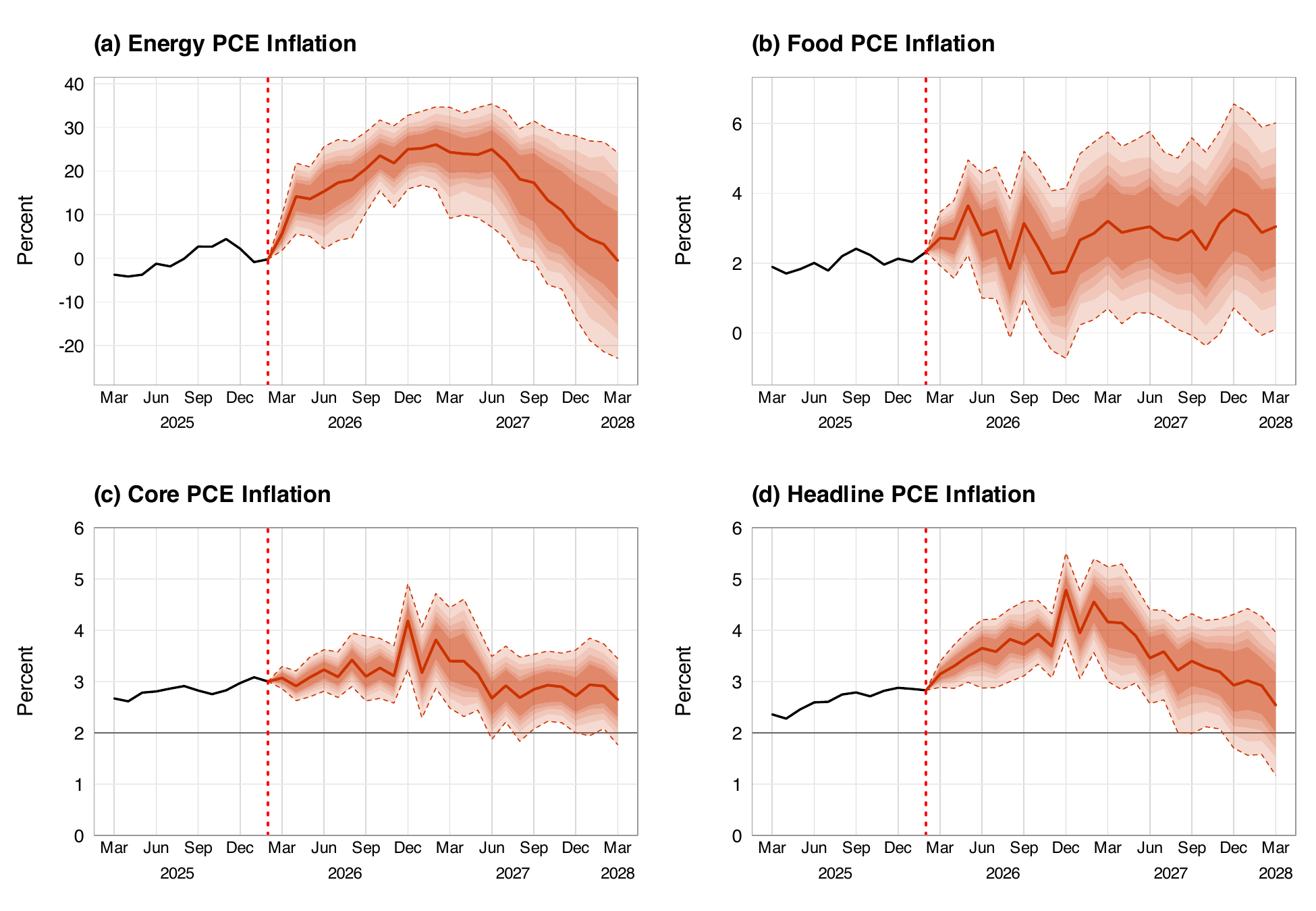}
  \caption*{\footnotesize \textbf{Notes}: The black line is realized inflation up to February 2026 which is indicated with the dotted
  line. The red line is the scenario-implied median forecast and the increasingly lighter shaded areas correspond to 50-60-70-80-90\% probability bands. The horizontal black line in panels (c) and (d) indicates the Fed's 2\% inflation target.}
  \label{fig:inflation}
  \vspace{-.6cm}
\end{figure}

\autoref{fig:inflation} displays the forecast distributions of the four inflation series conditional on the oil shortage scenario. Given its direct connection to oil prices, the energy component of PCE inflation, reported in panel (a), shows the greatest variability, with the median path rising steeply and exceeding 20\% shortly after inventories hit operational lows. Energy inflation remains above that threshold for about a year before gradually reverting to pre-war levels at the end of the forecast horizon. Increases in oil prices tend to raise the cost of raw agricultural products via the cost of farm fuels which might pass through to higher retail food prices \citep[see][]{BaumeisterKilian2014EP}. Panel (b) shows that predicted food price inflation rises sharply early in the conflict, reaching almost 4\%. Despite some volatility in 2026, it remains elevated at around 3\% until March 2028. Core PCE inflation stood at 3\% before the outbreak of the war and is conditionally predicted to reach 3.5\% by the summer of 2026, as can be seen in panel (c). By the end of the year, the median projection peaks at 4\%, after which it gradually declines, reaching pre-war levels by mid-2027. The scenario-implied median forecast of headline PCE inflation in panel (d) increases continuously over the course of 2026 from 3\% to 4\%, with a sharp peak of 4.8\% in December before slowly easing back to 3\% by the end of 2027 as the year-on-year oil impulse fades. It remains about one percentage point above the FOMC's inflation target through February 2028.

To quantify the upside risks to inflation resulting from the oil shortage scenario, \autoref{tab:inflrisk} reports for various forecast horizons the likelihood of headline (panel a) and core (panel b) PCE inflation exceeding the FOMC's inflation objective of 2\% and the pre-war level of 3\% as well as the percentage points (pp) by which inflation is expected to exceed those thresholds conditional on landing above them. Consistent with the predictive distributions in \autoref{fig:inflation}, it is virtually certain that both headline and core inflation surpass the Fed's inflation target for the first 12 months after the outbreak of the war. Conditional on crossing the 2\% threshold, headline and core inflation are expected to be respectively 2.5 pp and 1.8 pp above target after one year. Chances of exceeding the FOMC's objective remain around 90\% for both inflation measures also at longer horizons with probability-weighted excesses of around 1 pp. Even the 3\% threshold is highly likely to be breached by headline inflation at horizons of up to one year, with expected tail outcomes of 0.5 and 1.5 pp. The likelihood of this event occurring roughly halves at the two-year horizon, adding only 0.3 pp to headline PCE on a probability-weighted basis. Risks of core inflation being in excess of 3\% are relatively high after 12 months with an expected over-run of around 0.9 pp, but subside somewhat thereafter with probability-weighted excesses of 0.08 and 0.15 pp after 18 and 24 months, respectively.

\begin{table}[tbp]
\centering
\caption{\centering Real-Time Assessment of Inflation Risks as of February 2026}
\label{tab:inflrisk}
\begin{threeparttable}
\setlength{\tabcolsep}{4pt}
\begin{tabular}{lcccccc}
\toprule
 & \multicolumn{3}{c}{$\tau=2\%$} & \multicolumn{3}{c}{$\tau=3\%$} \\
\cmidrule(lr){2-4}\cmidrule(lr){5-7}
Horizon & $P^{M}(\pi>\tau)$ & \begin{tabular}[b]{@{}c@{}}$E[\pi-\tau\mid$\\$\pi>\tau]$\end{tabular} & $P^{M}\times E$ & $P^{M}(\pi>\tau)$ & \begin{tabular}[b]{@{}c@{}}$E[\pi-\tau\mid$\\$\pi>\tau]$\end{tabular} & $P^{M}\times E$ \\
\midrule
\multicolumn{7}{c}{\emph{Panel (a): Headline PCE Inflation}} \\
$h=\phantom{1}3$ (May~2026) & 1.00 & 1.49 & 1.49 & 0.94 & 0.54 & 0.50 \\
$h=\phantom{1}6$ (Aug~2026) & 1.00 & 1.79 & 1.79 & 0.95 & 0.85 & 0.81 \\
$h=12$ (Feb~2027) & 1.00 & 2.52 & 2.52 & 0.99 & 1.54 & 1.53 \\
$h=18$ (Aug~2027) & 0.95 & 1.24 & 1.18 & 0.63 & 0.58 & 0.36 \\
$h=24$ (Feb~2028) & 0.86 & 1.12 & 0.97 & 0.46 & 0.61 & 0.28 \\
\midrule
\multicolumn{7}{c}{\emph{Panel (b): Core PCE Inflation}} \\
$h=\phantom{1}3$ (May~2026) & 1.00 & 1.08 & 1.08 & 0.62 & 0.23 & 0.14 \\
$h=\phantom{1}6$ (Aug~2026) & 1.00 & 1.42 & 1.42 & 0.91 & 0.48 & 0.43 \\
$h=12$ (Feb~2027) & 1.00 & 1.80 & 1.80 & 0.93 & 0.88 & 0.82 \\
$h=18$ (Aug~2027) & 0.91 & 0.77 & 0.70 & 0.25 & 0.32 & 0.08 \\
$h=24$ (Feb~2028) & 0.97 & 0.94 & 0.91 & 0.42 & 0.37 & 0.15 \\
\bottomrule
\end{tabular}
\begin{tablenotes}\footnotesize
\item \textbf{Notes}: $P^{M}(\pi>\tau)$ denotes the marginal probability of exceeding $\tau$, $E[\pi-\tau\mid \pi>\tau]$ denotes the conditional expected excess in percentage points, and $P^{M}\times E$ is their product.
\end{tablenotes}
\end{threeparttable}
\end{table}

\subsubsection{Comparison with Survey Evidence}
To gauge the implications of our oil shortage scenario for the predictive distributions relative to survey-based assessments of inflationary developments, we compare our scenario-implied, model-based inflation forecast to publicly available forecasts from the Survey of Professional Forecasters (SPF) and the New York Fed's Survey of Market Expectations (SME) which were both taken around our February 2026 forecast origin. We use those surveys since they provide density forecasts in the form of histograms, for core PCE inflation in the case of the SPF and for headline PCE inflation in the case of the SME. Both report Q4/Q4 forecasts for 2026 and 2027, which we compare to our forecasts in November of the same years.

\begin{figure}[t]
\caption{\centering Comparison of Conditional Model-Based Density Forecasts with Survey-Based Histograms as of February 2026}
\vspace{-.3cm}
  \centering
  \includegraphics[width=0.98\textwidth]{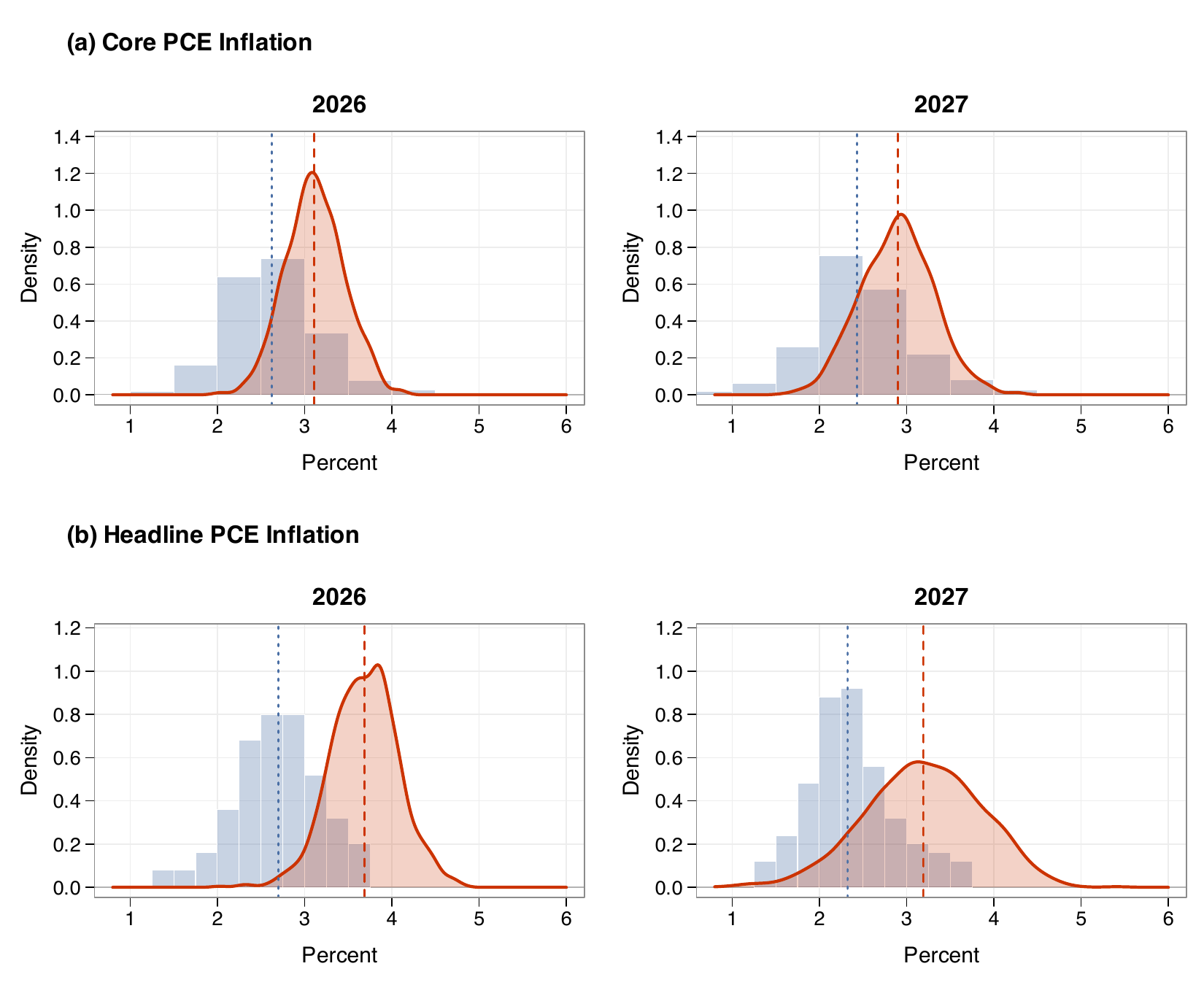}
  \caption*{\footnotesize \textbf{Notes}: The Q4/Q4 inflation rate is approximated by the mid-quarter (November) year-on-year forecasts generated by our monthly model. Survey histograms are converted to densities by dividing the bin probability by the bin width. The dashed red line indicates the model-based median and the dotted blue line the survey-based median.}
  \label{fig:survey}
  \vspace{-.6cm}
\end{figure}

As shown in \autoref{fig:survey}, the pre-conflict survey forecasts for 2026 and 2027 were centered at roughly 2.5\%, with more upside than downside risk to core inflation in 2027 and more balanced distributions in the other cases. The scenario-conditional densities lie well to the right of these pre-war survey densities, especially for headline PCE inflation. The conditional core inflation forecasts for 2026 and 2027 are centered around 3\%, whereas the median forecasts for headline inflation are 3.7\% in 2026 and 3.2\% in 2027. While the shapes of the predictive densities for both core and headline inflation for 2026 show more right-tail than left-tail skew, they look more symmetric for 2027, but have a longer upper than lower tail.

\subsection{Inflation Expectations Across Demographic Groups}

As noted previously, household inflation expectations are known to be particularly sensitive to changes in gasoline prices which are one of the most visible items in the consumption basket. But the way in which individual households form expectations about future inflation is far from uniform and depends on demographic characteristics. Using data from the University of Michigan survey, \cite{PfajfarSantoro2008} document substantial heterogeneity and asymmetries in the expectation formation process across different groups. They show that income, education, and gender matter for the level and accuracy of expectations as well as the speed and frequency at which survey respondents update their views. Recent evidence underscores that household inflation expectations differ systematically across demographic groups and respond strongly to energy and food price shocks \citep[e.g.,][]{DAcuntoetal2021, Dietrich2024, Hajdinietal2024, BaumeisterFrankHuberKoop2026}. It is therefore of interest to examine not only how aggregate inflation expectations evolve conditional on a war-induced surge in gasoline prices but also how adjustment differs across demographic groups. We focus on age, gender, and income tercile.

To quantify the conditional effects on inflation expectations across different households, we employ BART models analogous to the Phillips curve given in \eqref{eq:pcbart}. We replace actual inflation with one-year-ahead expected inflation from the Michigan survey and use the year-on-year growth rate of the retail gasoline price implied by the scenario rather than that of the crude oil price since the war-induced supply gap reaches households at the pump \citep[see, e.g.,][]{Binder2018, JoKlopack2025}. Given the mapping from Brent to gasoline prices via the rule of thumb in \eqref{eq:rot}, the information content is basically the same. We estimate separate models for the mean inflation expectations of each demographic group and construct forecasts conditional on the scenario-implied gasoline price.

\begin{figure}[t]
\caption{\centering Scenario-Implied Household Inflation Expectations}
\vspace{-0.2cm}
  \centering
  \includegraphics[width=0.99\textwidth]{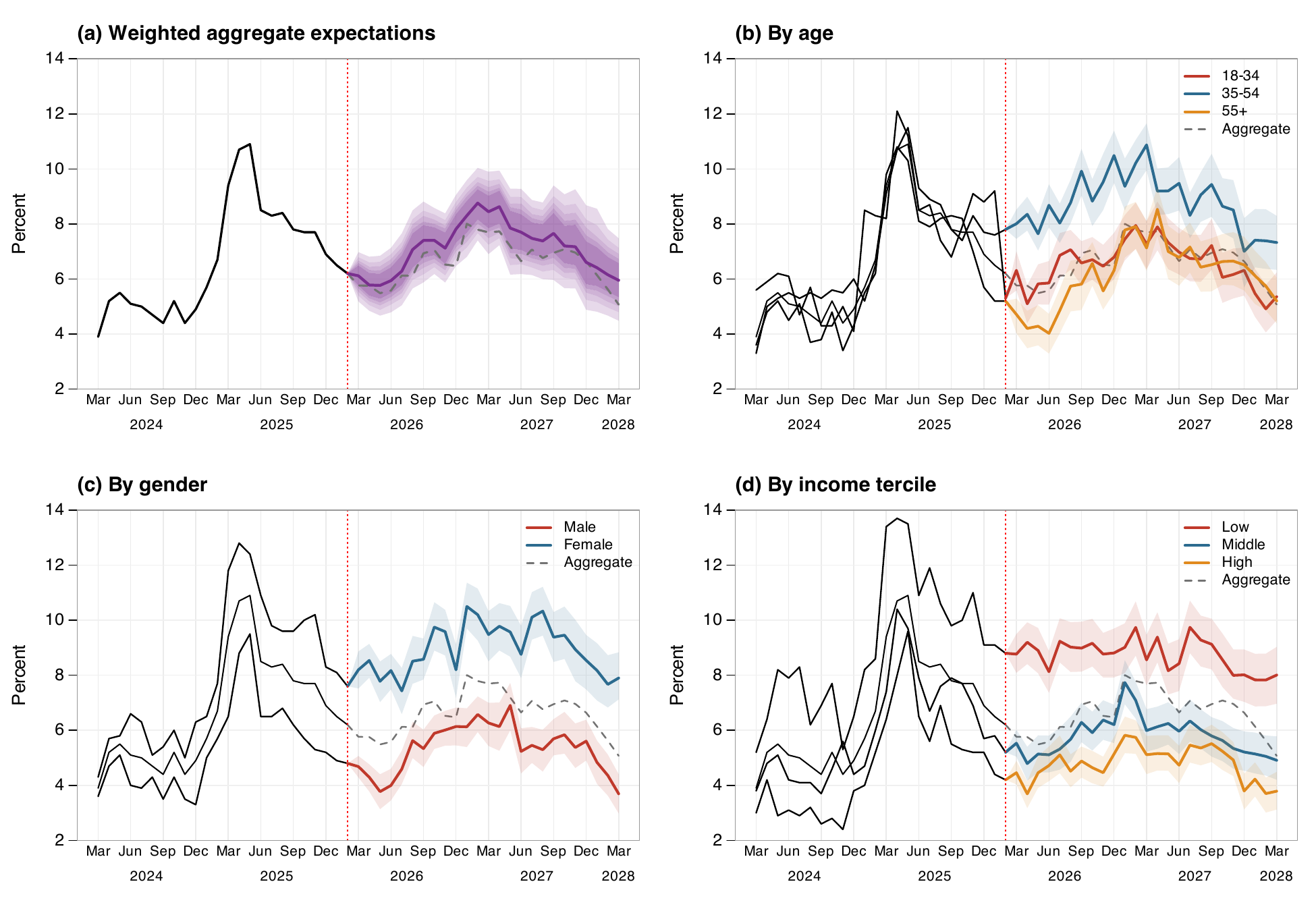}
  \caption*{\footnotesize \textbf{Notes}: Solid black lines show 24 observations of mean year-ahead household inflation expectations in the aggregate and by age, gender, and income tercile up to February 2026. The solid purple line in panel (a) is the median forecast aggregated from the subgroups. The gray dashed line is the aggregate expectation modeled directly and included in every panel for reference. Shaded areas in panel (a) correspond to 50-60-70-80-90\% probability bands, while in panels (b) to (d) only 25-75\% credible intervals are shown for readability.}
  \label{fig:expsub}
  \vspace{-.6cm}
\end{figure}

\autoref{fig:expsub} displays the predictive distributions of average year-ahead household inflation expectations conditional on the oil shortage scenario. Panel (a) shows the aggregate projection obtained by summing the conditional forecasts of the various demographic groups using estimated subgroup weights.\footnote{Because the survey does not publish the group shares, the weights are estimated so as to reconcile the subgroup means with the published aggregate series. Note that all series are survey means, which run above the median of year-ahead inflation expectations from the Michigan survey.} Starting from 6.2\% in February 2026, aggregate inflation expectations remain relatively stable in the spring of 2026 when emergency stock releases keep pump prices in check. Once inventories hit rock bottom and gasoline prices start climbing, expectations rise steadily and peak at 9\% in February 2027, with the 90\% probability band ranging from 7.5\% to 10\%. Thereafter, expectations gradually recede as the year-on-year increase in gasoline prices fades, returning to their pre-war level of about 6\% by March 2028. The bottom-up aggregate runs above the median expectations path estimated directly on the aggregate series, by roughly one percentage point on average. Because each group carries its own persistence and nonlinear response to the scenario-implied gasoline price path, the weighted sum of the group-level forecasts need not coincide with the forecast of the aggregate.

Panels (b) to (d) break these dynamics down by age, gender, and income tercile and reveal considerable heterogeneity. The well-documented differences across demographic groups persist throughout the forecast horizon. Inflation expectations are consistently higher among female and lower-income respondents than among their male and higher-income counterparts. Across age groups, survey participants aged 35 to 54 are the most sensitive, recording the highest expectations. The surge in pump prices raises expectations across all groups, with most of them peaking in early 2027 in line with the gasoline price path, but the size of the response differs markedly. Male respondents and those aged 55 and older seem the most optimistic that the massive inventory drawdown will keep inflation in check since their expectations first fall to about 4\% by mid-2026. But once the buffer is exhausted, their inflation expectations rise the most, to a peak of respectively 6.5\% and 8.5\% in April 2027. While female respondents and those aged 35 to 54 start at a higher level of around 8\% before the outbreak of the war, they increase their expectations more gradually by close to three percentage points peaking above 10\% in early 2027. The youngest cohort of survey participants frontloads their adjustment, raising inflation expectations by two percentage points within the first six months of the conflict. Overall, the gender gap widens from 2.8 percentage points before the war to 3.6 percentage points on average over the forecast horizon and age groups which displayed close comovement pre-crisis start to diverge notably.

The most striking asymmetry emerges across income groups. Low-income households have the highest inflation expectations of almost 9\% at the onset of the war but barely react to the scenario-induced gasoline price path, expecting inflation to increase by less than one percentage point at their peak in July 2027. In contrast, middle-income households raise their inflation expectations steeply by about 2.5 percentage points to 7.7\% in January 2027, while high-income households are more moderate and adjust their expectations by about 1.6 percentage points to 5.8\%. The conditional response paths imply a narrowing of the gap between low- and high-income households, from 4.6 percentage points before the war to 4 percentage points on average. The weaker response of low-income households is in keeping with evidence provided in \citet{CoibionGorodnichenko2015}. It is also consistent with the differences in information processing and reference prices across demographic groups documented by \citet{PfajfarSantoro2008}, which imply that the same oil supply shock need not propagate equally to the expectations of all households.

\section{Conclusions}

Motivated by broad interest in scenario analysis, the historical prevalence of oil price risks in monetary policy deliberations, and the ongoing conflict between the US and Iran, this paper develops a versatile and transparent framework for evaluating tail risks for the inflation outlook conditional on oil market scenarios. We conduct a forward-looking risk analysis where we assess the consequences of a tail event resulting from the war-induced rundown in global petroleum inventories after the blockade of the Strait of Hormuz in March 2026. This real-time risk assessment yields three main takeaways. First, the inventory buffer delays but does not prevent an oil price surge. Once petroleum stocks reach their historical minimum, the real Brent price and refining margins climb to new highs, while oil companies take their time to develop an appetite for investments in productive capacity. As a result of market tightness, the probability of oil prices surpassing \$200 rises to 28\%. Second, these risks pass through quickly to the pump, with the national average retail price of gasoline having a better-than-even chance of exceeding its 2022 record of \$5 per gallon a year after the outbreak of hostilities. Fuel price risks are skewed to the upside, unevenly distributed across the country, and highest on the West Coast. Third, our oil shortage scenario materially raises inflation risks. Headline and core PCE inflation peak close to 5\% and 4\%, respectively, with upward skewness in their predictive distributions. The probability of being at or below the Federal Reserve's 2\% inflation target two years after the closure of the Strait of Hormuz is 14\% for headline and 3\% for core inflation. Household inflation expectations rise by about 3 percentage points, with significant heterogeneity across demographic groups. Because these risks are asymmetric and state-dependent, predictive densities from linear models would greatly understate their likelihood. While our application focuses on scenarios for oil and gasoline prices and their implications for inflation dynamics associated with the current conflict in the Middle East, our proposed approach is more general. The three-component framework we develop can be applied to other sources of risk in global oil markets and other macroeconomic variables `at risk'.

\newpage

\appendix

\section{Data}\label{appx:data}
Table A1 provides details on the data and their sources used in the various parts of the paper. The three main sources for oil-related data are the Monthly Energy Review (MER), the International Energy Statistics (IES), and the Short-Term Energy Outlook (STEO) compiled by the US Energy Information Administration (EIA). The main source for US macroeconomic data is the Federal Reserve Economic Data (FRED) database maintained by the Federal Reserve Bank of St. Louis.
\setcounter{table}{0}
\renewcommand{\thetable}{A\arabic{table}}
\begin{table}[H]
\centering
\caption{\centering Variables and Sources}
\label{tab:data}
\begin{threeparttable}
\begin{tabular}{l >{\raggedright\arraybackslash}p{6.0cm}}
\toprule
\textbf{Variable} & \textbf{Source (mnemonic)} \\
\midrule
\vspace{0.2cm}
\textbf{Oil and Gasoline Market Analysis} \\
Spot price for Brent crude oil\tnote{a} & EIA \\
Global liquid fuels consumption\tnote{b} & STEO \\
OECD petroleum inventories\tnote{c} & IES \\
Global economic conditions indicator & \cite{BaumeisteretalRESTAT2022} (GECON) \\
CRB spot price index of raw industrial materials & Global Financial Data \\
Worldwide oil rig counts\tnote{d} & Baker-Hughes \\
\vspace{0.3cm}
US retail gasoline price\tnote{e} & MER (Table 9.4, column C) \\
\midrule
\vspace{0.2cm}
\textbf{Inflation Analysis} \\
Core PCE price index & FRED (PCEPILFE) \\
PCE price index for energy goods and services & FRED (DNRGRG3M086SBEA) \\
PCE price index for food & FRED (DFXARG3M086SBEA) \\
Output gap: log(real GDP/CBO potential)\tnote{f} & FRED (GDPC1 and GDPPOT) \\
U of Michigan year-ahead inflation expectations & FRED (MICH) \\
U of Michigan consumer sentiment index & FRED (UMCSENT) \\
Excess bond premium & \cite{FavaraEtAl2016} update of \cite{GilchristZakrajsek:2012:AER} (EBP) \\
\vspace{0.1cm}
Geopolitical risk & \cite{CaldaraIacoviello2022} (GPRH) \\
\bottomrule
\end{tabular}
\begin{tablenotes}
  \footnotesize\item[a] Since this series only starts in 1987.5, we extend it back to 1975.1 with the growth rate of the monthly refiner acquisition cost of imported crude oil (MER, Table 9.1). The Brent spot price is deflated with the US consumer price index for all urban consumers (FRED: CPIAUCSL).
  \footnotesize\item[b] Since this series only starts in 1990.1, we backcast it to 1982.1 with the growth rate of OECD refined petroleum products consumption (IES) and further back until 1975.1 with the growth rate of global crude oil production (IES). The series is seasonally adjusted with the X13-ARIMA procedure.
  \footnotesize\item[c] Since this series only starts in 1988.1, we extend the data back to 1975.1 with the growth rate of US total petroleum stocks (MER, Table 3.4).
  \footnotesize\item[d] Seasonally adjusted with the X13-ARIMA procedure.
  \footnotesize\item[e] Since the price for unleaded regular gasoline is only available from 1976.1 onward, we backcast it to 1973.1 with the growth rate of the price for leaded gasoline from the same source (column B).
  \footnotesize\item[f] The output gap is interpolated from quarterly to monthly frequency with the Chow-Lin approach \citep{ChowLin1971} using industrial production (FRED: INDPRO) as the higher-frequency indicator.
\end{tablenotes}
\end{threeparttable}
\end{table}

\clearpage
\section{Gasoline Pass-Through: Cross-Check}\label{app:ols}
The rule of thumb in \eqref{eq:rot} fixes the crude oil-to-gasoline slope at
2.5 cents per gallon for every dollar increase per barrel of oil. \autoref{fig:olsgas} cross-validates
this choice with an updated sample. A regression of the US retail gasoline price on an intercept and the cost of Brent crude over the period January 1973-February 2026 yields
\begin{equation*}
p^{gas} = \underset{(0.019)}{0.48} + \underset{(0.0004)}{0.0317}\times p^{oil},
\qquad R^{2}=0.92,
\end{equation*}
with standard errors in parentheses below the coefficient estimates. The slope is very precisely estimated. At \$0.032 it is
slightly steeper than the rule-of-thumb value, but over the price range relevant to the scenario the two lines are close.

\setcounter{figure}{0}
\renewcommand{\thefigure}{A\arabic{figure}}
\begin{figure}[H]
\caption{\centering Relationship between Oil and Gasoline Prices}
  \centering
  \includegraphics[width=0.9\textwidth]{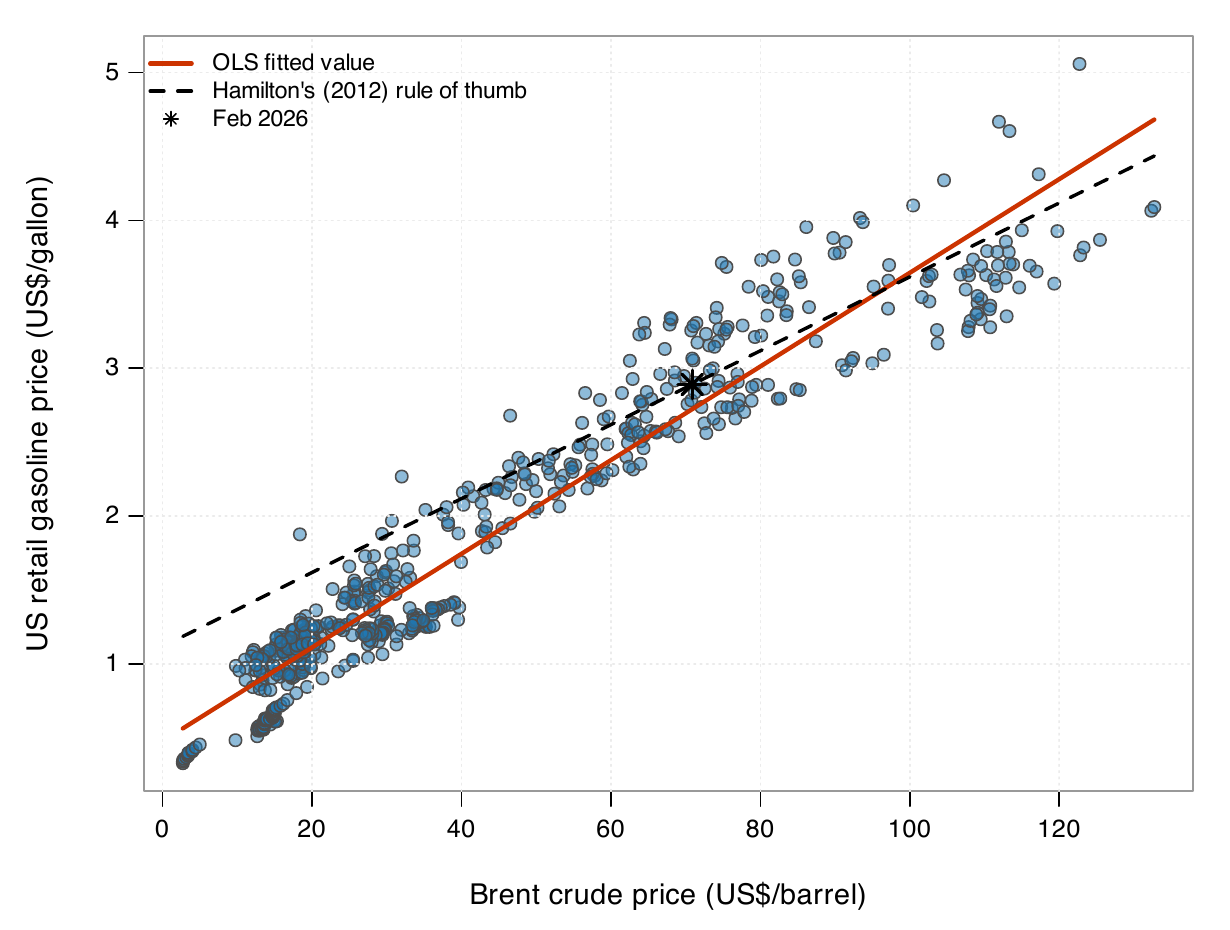}
  \label{fig:olsgas}
\end{figure}

\section{Predictive Evidence}

We assess the accuracy of tail forecasts for the real price of oil from the Bayesian VAR with stochastic volatility (BVAR-SV) and the BART model that mixes linear and nonlinear components and features stochastic volatility (mixBART-SV), against a random walk without drift, which has been identified as a tough benchmark to beat \citep[see, e.g.,][]{BaumeisteretalRESTAT2022}.  We also include a random walk with stochastic volatility (RW-SV) to gauge how much of the forecasting performance derives from allowing for heteroskedasticity.\footnote{All specifications use a random-walk law of motion for the log-volatilities.}

The models use the same variable set and lag order as in \autoref{sec:setup}. Our forecasting design is recursive: we produce $h$-step-ahead predictive distributions for $h = 1, 3, 6$, and $12$ months, extend the estimation sample by one month, re-estimate the models, and repeat. We use observed values from the May 2026 data vintage to assess the forecasting performance over the hold-out period 2020M1 to 2026M2. Since our focus is on tail forecasts, we evaluate the predictive distributions using quantile scores at $\tau = 0.05$ and $\tau = 0.95$, reported relative to the no-change benchmark.

\begin{table}[htbp]
\centering
\caption{\centering Tail Forecast Accuracy for the Real Price of Oil, 2020M1--2026M2}
\label{tab:tails2020}
\sisetup{table-format=1.3, table-space-text-post={$^{***}$}, table-align-text-after=false,
         reset-text-series=false, text-series-to-math=true, propagate-math-font=true}
\setlength{\tabcolsep}{12pt}
\begin{tabular}{l*{4}{S}}
\toprule
 & \multicolumn{4}{c}{Forecast horizon} \\
\cmidrule(lr){2-5}
 & {$h=1$} & {$h=3$} & {$h=6$} & {$h=12$} \\
\midrule
\multicolumn{5}{l}{\emph{Panel A: 5\% quantile}} \\[2pt]
mixBART-SV & \bfseries 0.804 & \bfseries 0.988 & 0.587$^{***}$ & 0.459$^{***}$ \\
BVAR-SV & 0.845 & 1.009 & \bfseries 0.583$^{***}$ & \bfseries 0.458$^{***}$ \\
RW-SV & 0.935 & 1.040 & 0.776$^{**}$ & 0.722$^{**}$ \\
\addlinespace
\multicolumn{5}{l}{\emph{Panel B: 95\% quantile}} \\[2pt]
mixBART-SV & \bfseries 0.652 & \bfseries 0.975 & \bfseries 0.897 & \bfseries 0.587$^{***}$ \\
BVAR-SV & 0.915 & 1.003 & 0.921 & 0.600$^{***}$ \\
RW-SV & 0.731 & 1.038 & 1.029 & 0.810 \\
\bottomrule
\end{tabular}
\caption*{\footnotesize \textbf{Notes}: Entries are average quantile scores relative to the no-change forecast of the
real price of Brent crude where values below one indicate greater accuracy than the benchmark. Boldface denotes the most accurate
model in each column. Asterisks denote rejection of equal predictive accuracy relative to
the no-change benchmark using the Diebold-Mariano test \citep{DieboldMariano1995} with Newey-West standard errors \citep{NeweyWest1987}
($h-1$ lags): $^{***}$ 1\%, $^{**}$ 5\%, $^{*}$ 10\%.}
\end{table}
\newpage
\bibliographystyle{frbcle}
\bibliography{lit}

\end{document}